\documentclass[aps,preprint,prd,showpacs,showkeys,nofootinbib,tightenlines]{revtex4-1}

\usepackage{graphicx}  
\usepackage{amsmath}
\usepackage{bm}  
\usepackage{ulem}
\usepackage{amssymb}
\usepackage{comment}
\usepackage{lineno}
\newcommand{\bea}{\begin{eqnarray}}
\newcommand{\eea}{\end{eqnarray}}
\newcommand{\beq}{\begin{equation}}
\newcommand{\eeq}{\end{equation}}
\newcommand{\bqa}{\begin{eqnarray}}
\newcommand{\eqa}{\end{eqnarray}}

\begin{document}

%%%%%%%%%%%%%%%%%%%%%%%%%%%%%%%%%%%%%%%%%%
\title{
Branching Fractions of the $\bm{X(3872)}$
}

\author{Eric Braaten}
\email{braaten.1@osu.edu}
\affiliation{Department of Physics,
         The Ohio State University, Columbus, OH\ 43210, USA}

\author{Li-Ping He}
\email{he.1011@buckeyemail.osu.edu}
\affiliation{Department of Physics,
         The Ohio State University, Columbus, OH\ 43210, USA}

\author{Kevin Ingles}
\email{ingles.27@buckeyemail.osu.edu}
\affiliation{Department of Physics,
         The Ohio State University, Columbus, OH\ 43210, USA}

\date{\today}
%\date{November 2007}

\begin{abstract}
The recoil momentum spectrum from the decay $B^+ \to K^+ +\mathrm{anything}$ has recently
been measured  by the BaBar collaboration. 
The spectrum  has a peak with invariant mass near  the mass of the $X(3872)$ meson.
The preliminary measurement by the BaBar collaboration implies that its branching fraction
 into $J/\psi\, \pi^+\pi^-$  is about 4\%.
We emphasize that this is the branching fraction for the entire resonance feature from $B^+$-to-$K^+$ transitions, 
which  includes a $D^{*0} \bar D^0$ and $D^0 \bar D^{*0}$  
threshold enhancement as well as a possible bound state  below the threshold or a virtual state.
If the  $X$ is a bound state  of charm mesons, its branching fraction into $J/\psi\, \pi^+\pi^-$ 
should be considerably larger than that of the $X$ resonance feature.
We use measurements of branching ratios of the $X$ to put an upper bound on this branching fraction of 33\%. 
We also constrain the parameters of the simplest plausible model for the line shapes of the $X$
using the precise measurement of the resonance energy of the $X$
and an estimate of the branching fraction into $D^0 \bar D^0 \pi^0$ and $D^0 \bar D^0\gamma$ from
the $X$ resonance feature from $B^+$-to-$K^+$ transitions.
\end{abstract}

\smallskip
\pacs{14.80.Va, 67.85.Bc, 31.15.bt}
\keywords{
Exotic hadrons, charm mesons.}
\maketitle
%%%%%%%%%%%%%%%%%%%%%%%%%%%%%%%%%%%%%%%%%%

%%%%%%%%%%%%%%%%%%%%%%%%%%%%%%%%%%%%%%%%%%
\section{Introduction}
\label{sec:Introduction}

%%%%%%%%%%%%%%%%%%%%%%%%%%%%%%%%%%%%%%%%%

Dozens of exotic heavy hadrons whose constituents include a heavy quark and its antiquark
have been discovered in high energy physics experiments since the beginning of the century
\cite{Chen:2016qju,Hosaka:2016pey,Lebed:2016hpi,Esposito:2016noz,Guo:2017jvc,Ali:2017jda,Olsen:2017bmm,Karliner:2017qhf,Yuan:2018inv,Brambilla:2019esw}. 
The first one to be discovered was the $X(3872)$ meson,
whose constituents include a charm quark and its antiquark.
It was discovered by the Belle Collaboration in 2003 
through its decay into $J/\psi\, \pi^+\pi^-$ \cite{Choi:2003ue}. 
Its $J^{PC}$ quantum numbers  were finally determined 
by the LHCb collaboration in 2013 to be $1^{++}$ \cite{Aaij:2013zoa}.
Its mass  is extremely close to the $D^{*0} \bar D^0$ scattering threshold,
with recent measurements indicating that the difference is at most about 0.2~MeV \cite{Tanabashi:2018oca}.
These facts suggest that $X(3872)$ is a weakly bound S-wave charm-meson molecule
with the flavor structure
%===============
\begin{equation}
\big| X(3872) \rangle = \frac{1}{\sqrt2}
\Big( \big| D^{*0} \bar D^0 \big\rangle +  \big| D^0 \bar D^{*0}  \big\rangle \Big).
\label{Xflavor}
\end{equation}
%===============
However, there are alternative models for the $X$ that have  not been excluded
\cite{Chen:2016qju,Hosaka:2016pey,Lebed:2016hpi,Esposito:2016noz,Guo:2017jvc,Ali:2017jda,Olsen:2017bmm,Karliner:2017qhf,Yuan:2018inv,Brambilla:2019esw}.
The $X$ has been observed in the {\it constituent decay modes} $D^0 \bar D^0 \pi^0$ and $D^0 \bar D^0 \gamma$,
which receive contributions from the decay of a constituent  $D^{*0}$ or $ \bar D^{*0} $.
It has also been observed in five {\it short-distance decay modes},  including  $J/\psi\, \pi^+\pi^-$,  
whose ultimate final states  include particles with momenta larger than the pion mass.
Despite all the information on its decays,
 there is as of yet no consensus on the nature  of the $X$.

There may be aspects of the production of $X$ that are more effective at discriminating
between models than its decays.
If $X$ is a weakly bound charm-meson molecule, its production can proceed by the creation
of a charm-meson pair $D^{*0} \bar D^0$ or $D^0 \bar D^{*0}$ 
at short distances of order $1/m_\pi$ or smaller, where $m_\pi$ is the pion mass,
followed  by the binding of the charm mesons into $X$ at longer distances.
The production of $X$ can also proceed by the creation of a charm-meson pair $D^* \bar{D}^*$ 
at short distances followed by the rescattering of the charm mesons into $X$ and a pion  
 \cite{Braaten:2018eov,Braaten:2019yua,Braaten:2019sxh}
or into $X$ and a photon  \cite{Dubynskiy:2006cj,Guo:2019qcn,Braaten:2019gfj}  at longer distances.
There are Feynman diagrams for such rescattering processes in which 
three charm mesons whose lines form a triangle can all be near their mass shells simultaneously.
A {\it triangle singularity} therefore produces a narrow peak 
in the $X \pi$ or $X \gamma$ invariant mass near the $D^* \bar{D}^*$ threshold.
Guo pointed out that any high-energy process that can create an S-wave $D^{*0} \bar{D}^{*0}$ pair at short distances 
will also produce $X \gamma$ with a narrow peak near the $D^{*0} \bar{D}^{*0}$ threshold 
due to a charm-meson triangle singularity \cite{Guo:2019qcn}. 
We had noted previously that rescattering of an S-wave $D^* \bar{D}^*$ pair produces a narrow peak 
in the $X\pi$ invariant mass near the $D^* \bar{D}^*$ threshold in hadron collisions \cite{Braaten:2018eov,Braaten:2019sxh}
and in $B$ meson decay into $KX\pi$  \cite{Braaten:2019yua}, without noting the connection to triangle singularities.
In Ref.~\cite{Braaten:2019gfj}, we calculated the cross section for electron-positron annihilation into $X \gamma$ 
near the $D^{*0} \bar{D}^{*0}$ threshold, where rescattering of a P-wave $D^{*0} \bar{D}^{*0}$ pair 
produces a narrow peak due to a triangle singularity.
The peaks in the $X \pi$ or $X \gamma$ invariant mass from charm-meson triangle singularities 
provide smoking guns for the identification of the $X$ as a charm-meson molecule.

It is important to have quantitative predictions for the height, width, and shape of the peaks from the 
charm-meson triangle singularities.  If the $X$ is a weakly bound charm-meson molecule,
the height of the peak in a specific decay  mode is controlled by the 
binding energy of the $X$ and by its branching fraction into that decay mode.
The BaBar collaboration has recently determined the inclusive branching fraction of the $B^+$ meson  into 
$K^+$ plus the $X$ resonance by measuring  the recoil momentum spectrum of the $K^+$ \cite{Wormser}.
The preliminary result implies that the branching fraction of the entire $X$ resonance feature 
 into $J/\psi\, \pi^+\pi^-$  is about 4\%.  
 This small branching fraction may suggest that it could be difficult to observe the peak 
 from a charm-meson triangle singularity in the $J/\psi\, \pi^+\pi^-$ decay mode.
 However the $X$ resonance feature includes 
 a threshold enhancement in the production of $D^{*0} \bar D^0$ and $D^0 \bar D^{*0}$
 as well as a possible narrow peak from the $X$ bound state.  We emphasize in this paper that
the branching fraction of the $X$ bound state into $J/\psi\, \pi^+\pi^-$
should be considerably larger than the corresponding branching fraction of the $X$ resonance feature.  

We begin in Section~\ref{sec:short-distance} by explaining why branching fractions 
of a bound state must be the same for all short-distance production mechanisms.
In Section~\ref{sec:universal}, we describe
the simplest plausible model for the line shapes of the $X$ resonance.
In Section~\ref{sec:Eresonance}, we discuss the short-distance production of the $X$  resonance 
and present a simple  theoretical prescription for the resonance energy.
We determine the resonance energy analytically for the simplest  model for the $X$ line shapes in three limits.
In Section~\ref{sec:Branching}, we discuss branching fractions for the $X$ resonance feature and for the $X$ bound state.
We use previous experimental results to estimate the branching fraction for the $X$ resonance feature 
from $B^+$-to-$K^+$ transitions into short-distance decay modes.
We  use our estimate for that branching fraction combined with the current result for the $X$ resonance energy 
to constrain  the parameters of the simplest model for the $X$ line shapes.
We summarize our results in Section~\ref{sec:Summary}.

%\newpage

%%%%%%%%%%%%%%%%%%%%%%%%%%%%%%%%%%%%
\section{Factorization in Short-distance Production}
\label{sec:short-distance}
%%%%%%%%%%%%%%%%%%%%%%%%%%%%%%%%%%%%

We consider pairs of particles with coupled S-wave scattering channels that we label by an index $i$
and that have nearby scattering thresholds.
We are particularly interested in the case of an S-wave resonance near the scattering threshold
 for the lowest channel, which we take to be at the energy $E=0$.
The transition rates between channels can be expressed in terms of scattering amplitudes $f_{ij}(E)$
that depend on the center-of-mass energy $E$.
The optical theorem for the scattering amplitudes that follows from the unitarity of the S-matrix is
%===========
\begin{equation}
2\, \mathrm{Im}\big[f_{ij}(E)\big] = \sum_k f_{ik}(E) \,f_{jk}(E)^*.
\label{fij:unitarity}
\end{equation}
%=====

If there is a resonance, the scattering amplitudes for all the coupled channels 
have a pole at the same complex energy $E_X - i \Gamma_X/2$, where $E_X$ and $\Gamma_X$ are real.
At complex energies $E$ near the pole, the scattering amplitudes can be expressed in the factored form
%===========
\begin{equation}
f_{ij}(E) \approx  \frac{- c_i \, c_j}{E - E_X + i \Gamma_X/2},
\label{fij-E:pole}
\end{equation}
%=====
where the energy-independent constants $c_i$'s are required  to be real by time-reversal symmetry.
If the pole is on the physical sheet of the complex energy $E$, the resonance is referred to as a {\it bound state}.
If the pole is on a different sheet of $E$, the resonance is referred to as a {\it virtual state}.
A {\it narrow bound state}  is one for which the pole energy satisfies
$E_X < 0$ and $\Gamma_X > 0$ with $|E_X|$ significantly larger than $\Gamma_X/2$.
In this case, $|E_X|$  can be interpreted as the {\it binding energy} and 
$\Gamma_X$ can be interpreted as the {\it decay width}  of the bound state.
Only in the case of a narrow bound state are the expressions for the scattering amplitudes 
near the pole in Eq.~\eqref{fij-E:pole} good approximations over a real range of the energy $E$.

There may be short-distance production mechanisms for the pairs of particles
that involve momentum scales much larger than those provided by $E_X$, $\Gamma_X$, and 
the energy differences between scattering thresholds.
The amplitude for the short-distance production of a pair of particles in the scattering channel $k$ 
can be expressed in the factored form $\sum_iB_i\, f_{ik}(E)$,
where the short-distance factor $B_i$ is insensitive to the energy $E$.
In the case of S-wave production channels, the $B_i$'s are constants.
Different short-distance production mechanisms will have different factors $B_i$.
The inclusive production rate summed over channels $k$ can be expressed as
 %===========
\begin{equation}
\frac{dR}{dE} = \sum_k \Big( \sum_i B_i \, f_{ik}(E) \Big)\, \Big(\sum_j B_j \, f_{jk}(E) \Big)^*.
\label{dR/dE:inclusive}
\end{equation}
%=====
The optical theorem in Eq.~\eqref{fij:unitarity} implies that the  inclusive production rate 
 can be expressed in the factored form
%===========
\begin{equation}
\frac{dR}{dE} = 2 \sum_{ij} B_i \, \mathrm{Im}\big[f_{ij}(E)\big] \, {B_j}^*.
\label{dR/dE:optical}
\end{equation}
%=====
The S-wave resonance produces an enhancement in the inclusive production rate 
near the threshold.  We refer to the entire resonantly enhanced contribution  as the {\it resonance feature}.
The energy dependence of $dR/dE$ defines the {\it inclusive line shape}.
A unique feature of a near-threshold S-wave resonance is that the resonance feature includes a peak 
in the production rate of pairs of particles just above the threshold, which we refer to as a {\it threshold enhancement}.
The resonance feature may also include contributions below the threshold from a bound state or a virtual state.

In the case of a narrow bound state, the inclusive production rate has a narrow peak
below the threshold with a maximum near $E_X$ and a width in $E$ of about $\Gamma_X$.
The pole approximation for the scattering amplitude in Eq.~\eqref{fij-E:pole}
can be used to express the inclusive production rate in Eq.~\eqref{dR/dE:optical}
at real energies $E$ near the peak  in a factored form:
%===========
\begin{equation}
\frac{dR}{dE} \approx
 \Big( \sum_i B_i c_i \Big) \frac{\Gamma_X}{(E - E_X)^2 + \Gamma_X^2/4}  \Big( \sum_j B_j c_j \Big)^*.
\label{dR/dE:factor}
\end{equation}
%=====
The  decay width $\Gamma_X$ in the numerator can be expanded 
as a sum of the partial widths of all the decay modes $k$  of the bound state: $\Gamma_X = \sum_k \Gamma_{X \to k}$.
The production rate $dR_k/dE$ in a specific decay mode $k$ can be expressed in the same factored form
in Eq.~\eqref{dR/dE:factor} with $\Gamma_X$ in the numerator replaced by  $\Gamma_{X \to k}$.
The branching fraction $\Gamma_{X \to k}/\Gamma_X$ can also be expressed as the ratio of an integral  of 
$dR_k/dE$ over an integral of $dR/dE$, where the integrals are over the narrow peak from the bound state. 
The factorized form of Eq.~\eqref{dR/dE:factor} guarantees that such a branching fraction
is independent of the production mechanism.

The independence of the branching fractions on the production mechanism is guaranteed only in the case 
of a narrow bound state and only for branching fractions obtained by integrating over the narrow bound-state peak.
In the case of a virtual state or a bound state that is not narrow, 
a branching fraction defined by a ratio of integrals should be expected to depend on the production mechanism.
Even in the case of a narrow bound state, if the integration region is extended to include the threshold enhancement,
a branching fraction defined by a ratio of integrals should be expected to
depend on the production mechanism.

%%%%%%%%%%%%%%%%%%%%%%%%%%%%%%%%%%%%
\section{Simplest Model for $\bm{X}$ Line Shapes} 
\label{sec:universal}
%%%%%%%%%%%%%%%%%%%%%%%%%%%%%%%%%%%%

In the case of the  $X(3872)$ resonance, the particles are charm mesons.
We denote the masses of $D^{*0}$ and $D^0$ by $M_{*0}$ and $M_0$, respectively.
We denote the masses of $D^{*+}$ and $D^+$ by $M_{*1}$ and $M_1$, respectively.
The reduced mass of $D^{*0} \bar D^0$ is  
$\mu=M_{*0}M_0/(M_{*0} \!+\!M_0)= 967$~MeV.
The difference between the scattering thresholds for $D^{*+} D^-$ and $D^{*0} \bar D^0$ is 
$\delta = (M_{*1}\!+\!M_1) - (M_{*0}\!+\!M_0)= 8.2$~MeV.
The decay  width of the $D^{*0}$ can be predicted from measurements of $D^*$ decays: 
$\Gamma_{*0} = (55.9 \pm 1.6)$~keV  \cite{Rosner:2013sha}. 
The corresponding momentum scale is $\sqrt{\mu \Gamma_{*0}} = 7.4$~MeV.
The present value of the difference $E_X$ between the mass of the $X$ 
and the energy of the $D^{*0} \bar D^0$ scattering threshold is \cite{Tanabashi:2018oca}
%===============
\begin{equation}
E_X \equiv M_X - (M_{*0}\!+\!M_0) =( +0.01 \pm 0.18)~\mathrm{MeV}.
\label{EX-exp}
\end{equation}
%===============
This value has been obtained from measurements in the $J/\psi\, \pi^+\pi^-$ decay mode.
The central value of $E_X$ is essentially at the $D^{*0} \bar D^0$ scattering threshold.
The value of $E_X$  lower by $1\sigma$ corresponds to a bound state with binding energy $|E_X| =0.17$~MeV. 

The line shape for the $X(3872)$  in the $D^0 \bar D^0 \pi^0$ decay mode
from the decay $B \to K(D^0 \bar D^0 \pi^0)$ has been measured by the Belle collaboration  \cite{Gokhroo:2006bt}.
The distribution in the energy $E$ defined by the difference between the $D^0 \bar D^0 \pi^0$ invariant mass
and the $D^{*0} \bar D^0$ threshold has a peak at 
$E=4.2^{+0.8}_{-1.7}$~MeV.\footnote{This result is obtained from
the peak position $M_{D^0\bar D^0\pi^0}$ from Ref.~\cite{Gokhroo:2006bt}, 
the mass $M_{D^0}$ from the PDG in 2006, 
and the mass difference $M_{D^{*0}}-M_{D^0}$ from the PDG in 2018 \cite{Tanabashi:2018oca}.
The errors have been added in quadrature.}
The fitted energy distribution decreases to a local minimum near 10~MeV before increasing.
The distribution up to that minimum can be identified with the $X$ resonance feature in the 
$D^0 \bar D^0 \pi^0$ decay mode.
The resolution was  insufficient to resolve any further substructure in the  resonance feature.

There have been many previous theoretical studies of the line shapes of the $X(3872)$ resonance.
The earliest such studies were carried out by Voloshin  \cite{Voloshin:2003nt,Voloshin:2007hh}.
The simplest analytic model for the line shapes considers only the single resonant channel 
with a neutral-charm-meson pair \cite{Braaten:2007dw}.
More elaborate analytic models take into account the coupling to a charmonium state 
with quantum numbers $1^{++}$  that  can be identified with the $\chi_{c1}(2P)$
 \cite{Hanhart:2007yq,Zhang:2009bv,Kalashnikova:2009gt,Artoisenet:2010va},
the coupling to a charged-charm-meson pair \cite{Braaten:2007ft,Artoisenet:2010va,Hanhart:2011jz,Kang:2016jxw},
and the coupling to the $J/{\psi} \, \omega$ channel  \cite{Braaten:2013poa}.
There have also been efforts to take into account the $D \bar D \pi$ channels by using an effective field theory
for charm mesons and pions called XEFT \cite{Fleming:2007rp,Braaten:2015tga}
or by solving Lippmann-Schwinger equations numerically \cite{Baru:2011rs,Schmidt:2018vvl}.

%%%%%%%%%%%%%%%%%%%%%%%%%%%%%%%%%%%%%%%%%%%%%%%%
\begin{figure}[ht]
\includegraphics*[width=0.8\linewidth]{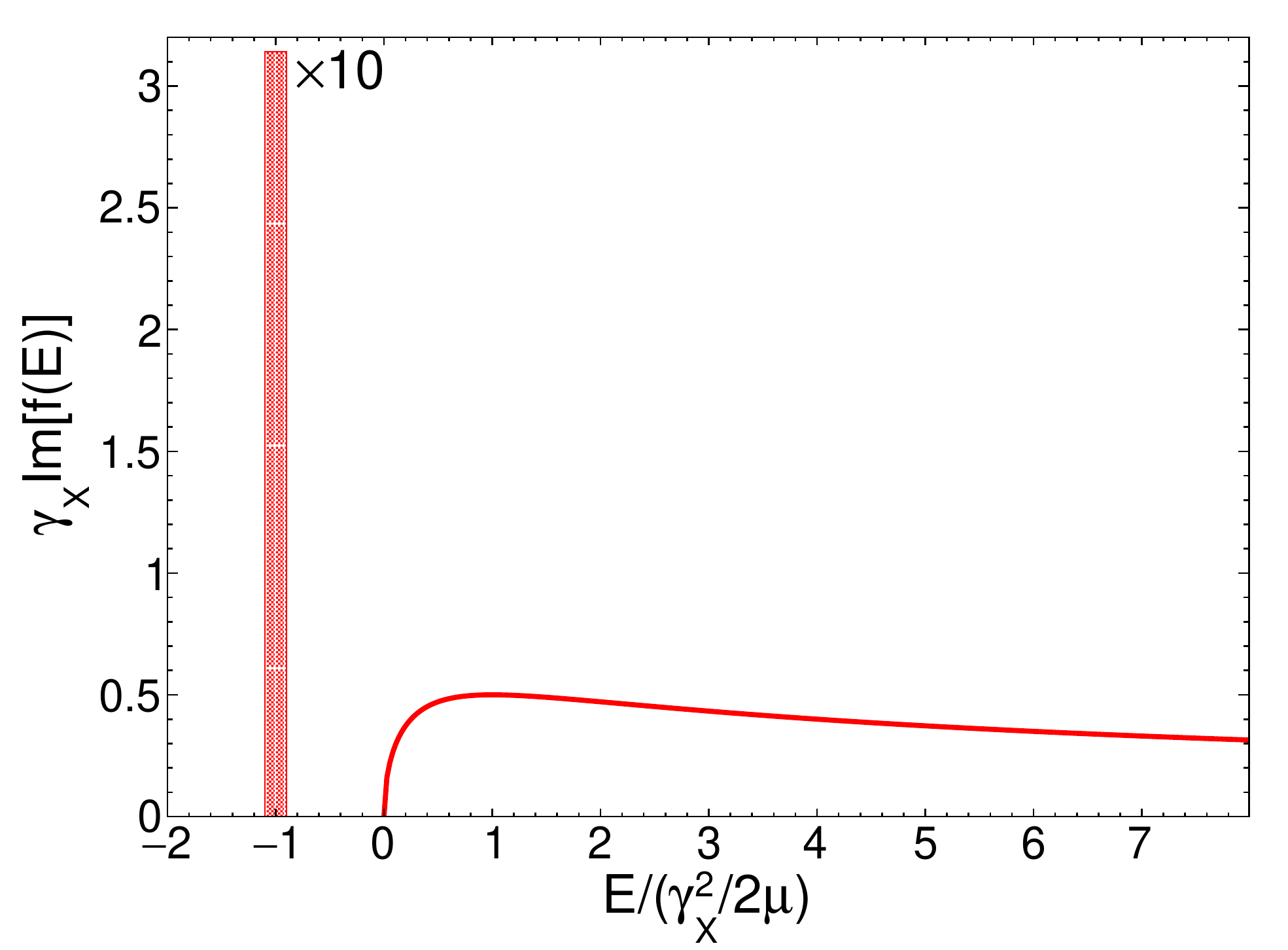} 
\caption{
Line shape $\mathrm{Im}[f_X(E+ i \epsilon)]$ in Eq.~\eqref{Imf-E} as a function of the energy $E$.
The resonance feature includes a threshold enhancement from production of the pair of particles 
above the scattering threshold at $E=0$.
If $\gamma_X >0$, the resonance feature also includes a delta function from production of the bound state 
below the scattering threshold. 
It is represented by the tall rectangle whose area should be multiplied by 10.
}
\label{fig:DstarDbar}
%\vspace*{0.0cm}
\end{figure}
%%%%%%%%%%%%%%%%%%%%%%%%%%%%%%%%%%%%%%%%%%%%%%

For particles with short-range interactions that produce an S-wave resonance sufficiently close to the scattering threshold, 
the scattering amplitude at very low energy has the simple universal form \cite{Braaten:2004rn}
%===========
\begin{equation}
f(E) = 
   \frac{1}{- \gamma_X  + \sqrt{-2\mu (E + i \epsilon)}},
\label{f-E}
\end{equation}
%===========
where $\mu$ is the reduced mass of the pair of particles
and $E$ is their total energy relative to the scattering threshold.  
Exact unitarity requires the inverse scattering length $\gamma_X$ to be real.
The  inclusive line shape is
%===============
\begin{subequations}
\begin{eqnarray}
\mathrm{Im}[f_X(E+ i \epsilon)] 
&=& \frac{\pi \gamma_X}{\mu} \delta(E + \gamma_X^2/2\mu) \, \theta(\gamma_X) \qquad (E<0),
\label{Imf-E-}
\\
&=& \frac{\sqrt{2\mu E}}{\gamma_X^2+2\mu E}  \qquad  \qquad \qquad  \qquad~ (E>0).
\label{Imf-E+}
\end{eqnarray}
\label{Imf-E}%
\end{subequations}
%===============
Eq.~\eqref{Imf-E+} implies that the maximum of the threshold enhancement is at $E = + \gamma_X^2/2\mu$.  
Eq.~\eqref{Imf-E-} implies that if $\gamma_X>0$, there is also a delta function 
at the negative energy $E_X=-\gamma_X^2/2\mu$ from production of the bound state.  
The threshold enhancement together with the delta function (if there is one) 
forms the {\it resonance feature}, which is illustrated in Fig.~\ref{fig:DstarDbar}.
Note that the integral of the line shape in Eq.~\eqref{Imf-E+} up to an energy $E_\mathrm{max}$ 
increases as $E_\mathrm{max}^{1/2}$. 
In contrast to the line shape of a Breit-Wigner resonance, the integral diverges as $E_\mathrm{max} \to \infty$.
This behavior introduces complications in the definitions of some resonance properties.

In the case of the  $X(3872)$, the superposition of neutral charm-meson pairs in Eq.~\eqref{Xflavor}
has a resonance in  the S-wave $1^{++}$ channel.
The nearest threshold for a coupled channel is that for the charged-charm meson pairs,
which is higher by 8.2~MeV.
The simplest plausible model for the resonant scattering amplitude $f(E)$ can be obtained 
from the universal amplitude in Eq.~\eqref{f-E} by making two changes  \cite{Braaten:2007dw}:
\begin{itemize}
\item
The effects of the width of the $D^{*0}$ are taken into account 
by replacing $E+ i \epsilon$ by  $E + i \Gamma_{*0}/2$.
\item
The effects of short-distance decay modes are taken into account 
by allowing the real parameter $\gamma_X$ to have a positive imaginary part.  
\end{itemize}
The resulting scattering amplitude is
%===========
\begin{equation}
f(E) =   \frac{1}{- \gamma_X  + \sqrt{-2\mu (E + i \Gamma_{*0}/2)}}.
\label{Amp-E}
\end{equation}
%===========
The only undetermined parameter in the amplitude is the complex inverse scattering length $\gamma_X$.  
The real and imaginary parts of $\gamma_X$ are both determined 
by the physics at short distances.  However $\mathrm{Re}[\gamma_X]$ 
is sensitive to the fine tuning of the physics at short distances.
For the purpose of order-of-magnitude estimates, 
we will assume $\mathrm{Im}[\gamma_X]$ is order $\sqrt{\mu \Gamma_{*0}}$.
The real part of  $\gamma_X$ can range from positive values much larger than $\sqrt{\mu \Gamma_{*0}}$
 to negative values with absolute value much larger than $\sqrt{\mu \Gamma_{*0}}$.

%%%%%%%%%%%%%%%%%%%%%%%%%%%%%%%%%%%%%%%%%%%%%%%%
\begin{figure}[t]
\includegraphics*[width=0.6\linewidth]{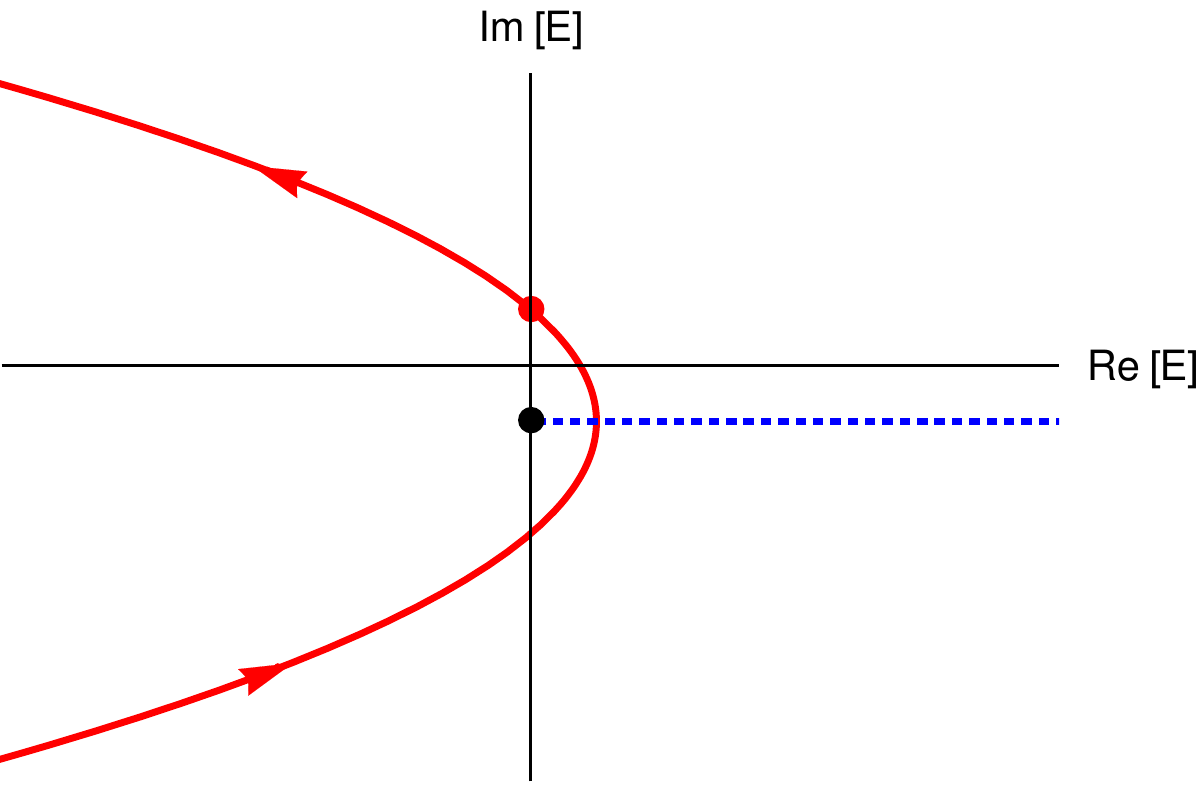} 
\caption{Analytic features of the  scattering amplitude  $f(E)$ in Eq.~\eqref{Amp-E}
in the  plane of the complex energy $E$.
The solid curve is the path of the pole as $\mathrm{Re}[\gamma_X]$ is varied with $\mathrm{Im}[\gamma_X]$ fixed.
 The arrows indicate the direction of increasing resonance energy $E_X$.
The branch point at $- i \Gamma_{*0}/2$ for the square-root branch cut (dashed line) is marked by a dot.
The dot on the solid curve marks the point on the second sheet where there is a zero-energy resonance with
$E_X=0$. We have set $\mathrm{Im}[\gamma_X] = \sqrt{\mu \Gamma_{*0}}$.}
\label{fig:Epolepath}
\end{figure}
%%%%%%%%%%%%%%%%%%%%%%%%%%%%%%%%%%%%%%%%%%%%%%

The scattering amplitude in Eq.~\eqref{Amp-E} is an  analytic  function of  the complex energy $E$
with a square-root branch point at $E= - i \Gamma_{*0}/2$.
We choose the branch cut to be along the line $\mathrm{Re}[E]>0$,  $\mathrm{Im}[E] = - i \Gamma_{*0}/2$.
The amplitude also has a pole at the energy $E_\mathrm{pole} = -\gamma_X^2/(2\mu)- i \Gamma_{*0}/2$. 
Its expression in terms of the real and imaginary parts of $\gamma_X$ is
%===========
\begin{equation}
E_\mathrm{pole} = - \frac{\mathrm{Re}[\gamma_X]^2 - \mathrm{Im}[\gamma_X]^2}{2 \mu}
- \frac{i}{2} \left( \Gamma_{*0} + \frac{2\, \mathrm{Re}[\gamma_X]\,\mathrm{Im}[\gamma_X]}{\mu} \right).
\label{Epole}
\end{equation}
%===========
The path of the pole in the plane of the complex energy $E$ as $\mathrm{Re}[\gamma_X]$ 
decreases with $\mathrm{Im}[\gamma_X]$ fixed is illustrated in Fig.~\ref{fig:Epolepath}.
The pole crosses the branch cut when $\mathrm{Re}[\gamma_X]=0$.
If $\mathrm{Re}[\gamma_X]>0$, the pole is on the physical sheet and the resonance is a bound state.
If $\mathrm{Re}[\gamma_X]<0$, the pole is on the second sheet and the resonance is a virtual state.
For complex energies $E$ near the pole, the scattering amplitude can be approximated by
%===========
\begin{equation}
f(E) \approx  \frac{- \gamma_X/\mu}{E - E_\mathrm{pole}},
\label{f-E:pole}
\end{equation}
%=====

The inclusive production rate of the resonance through a mechanism that creates the 
constituents at short distances is proportional to $\mathrm{Im}[f(E)]$.
The imaginary part of the scattering amplitude in Eq.~\eqref{Amp-E} at a real energy $E$ can be expressed as
%===========
\begin{equation}
\mathrm{Im} \big[f(E) \big] = 
\big|f(E) \big|^2
\left(  \mathrm{Im}[\gamma_X]  + \Big[\mu \sqrt{E^2+  \Gamma_{*0}^2/4} + \mu E \Big]^{1/2} \right).
\label{ImAmp-E}
\end{equation}
%===========
The unitarity condition $\mathrm{Im} [f(E)] \ge 0$ requires $\mathrm{Im}[\gamma_X]\ge 0$.
The term in Eq.~\eqref{ImAmp-E} with the factor $\mathrm{Im}[\gamma_X]$ can be interpreted as the contribution 
from the {\it short-distance decay} (SDD) channel, which consists of the decay modes
whose ultimate final states include particles with momenta larger than $m_\pi$.
The simplest model for the $X$ line shapes predicts that all the decay modes in the SDD channel 
have the same line shape proportional to $|f(E)|^2$.
In the case of the $X$ resonance, $\mathrm{Im}[\gamma_X]$ can be expressed as the sum of positive contributions 
from $J/\psi\, \pi^+\pi^-$ and all the other decay modes in the SDD  channel.
There may also be small short-distance contributions 
from the $D^0 \bar D^0 \pi^0$ and $D^0 \bar D^0 \gamma$ decay modes.
The other term in Eq.~\eqref{ImAmp-E} (which has the factor in square brackets raised to the power $\tfrac12$)
can be interpreted as the contribution from the {\it constituent decay} (CD) channel, 
which consists of the decay modes whose ultimate final states are $D^0 \bar D^0 \pi^0$ and $D^0 \bar D^0 \gamma$.  
This simplest model  for the $X$ line shapes predicts that the line shapes in the 
$D^0 \bar D^0 \pi^0$ and $D^0 \bar D^0 \gamma$ decay modes
differ only by multiplicative constants whose ratio is the branching ratio of $D^{*0}$ into $D^0\pi^0$ over $D^0\gamma$.
The line shape for the CD channel has a threshold enhancement 
from the production of $D^{*0} \bar D^0$ or $D^0 \bar D^{*0}$ above their scattering  threshold 
as well as contributions from the production of $D^0 \bar D^0 \pi^0$ or $D^0 \bar D^0 \gamma$ below that threshold.

Some aspects of a resonance feature are most conveniently quantified in terms of integrals of  line shapes over the energy.
If the inclusive line shape in Eq.~\eqref{ImAmp-E} is integrated over the energy from
$E_\mathrm{min}$ to $E_\mathrm{max}$, the integral diverges as $E_\mathrm{min} \to -\infty$ 
and as $E_\mathrm{max} \to +\infty$.  It is therefore necessary to introduce an additional prescription 
for the resonance defined by the scattering amplitude in Eq.~\eqref{Amp-E}.
A simple prescription for  the resonance is to declare it to be the energy range 
between a specified energy $E_\mathrm{min}$ below the scattering threshold 
and a specified energy $E_\mathrm{max}$ above the threshold.
Equivalently, we could declare the line shape to be given by Eq.~\eqref{ImAmp-E} for 
$E_\mathrm{min} < E < E_\mathrm{max}$ and to be zero outside that interval.
The sudden drops of the inclusive line shape from the function in Eq.~\eqref{ImAmp-E}  
to 0 below $E_\mathrm{min}$ and above $E_\mathrm{max}$ 
emphasizes the crudeness of this model for the threshold enhancement.
An alternative prescription for the resonance would be to smear the line shape in Eq.~\eqref{ImAmp-E} 
using a Gaussian function of $E$  whose width could be chosen to mimic the experimental energy resolution.
We choose to use the simpler prescription $E_\mathrm{min} < E < E_\mathrm{max}$ for the resonance,
because it can be used to obtain analytic results for some properties of the resonance in certain limits.

In order to make quantitative predictions for properties of the resonance,
we need to choose numerical values for  $E_\mathrm{min}$ and $E_\mathrm{max}$.
The energy range from $E_\mathrm{min}$ to $E_\mathrm{max}$ must be wide enough to cover most of the resonance.  
The values of  $E_\mathrm{max}$ must be at most  8.2~MeV 
to avoid complications from the coupling to the charged-charm-meson-pair channel.
The values of $|E_\mathrm{min}|$ and $E_\mathrm{max}$ must be less than $m_\pi^2/M_D \approx 10$~MeV to 
avoid contributions involving the large momentum scale $m_\pi$.
We choose $E_\mathrm{min}$ and $E_\mathrm{max}$ rather arbitrarily to be the nearest relevant kinematic thresholds.
We choose $E_\mathrm{min}$ to be the  $D^0 \bar D^0 \pi^0$ threshold: $E_\mathrm{min}= -7.0$~MeV.
We choose $E_\mathrm{max}$ to be the $D^{*+} D^-$ threshold: $E_\mathrm{max}= +8.2$~MeV.

%\newpage

%%%%%%%%%%%%%%%%%%%%%%%%%%%%%%%%%%%%
\section{Resonance energy}
\label{sec:Eresonance}
%%%%%%%%%%%%%%%%%%%%%%%%%%%%%%%%%%%%

The difference between the line shape of a near-threshold S-wave resonance 
and that of a conventional resonance 
complicates the definition of the $X$ resonance energy.
In this Section, we introduce a theoretical prescription for the resonance energy $E_X$.
In the case of the simplest model for the $X$ lines shapes, we determine $E_X$ analytically in three  limits.

A possible theoretical definition of a resonance energy,  such as $E_X$ in Eq.~\eqref{EX-exp},
 is the resonance-weighted average of the energy,
which can be defined mathematically as a ratio of integrals.
Such a definition is not applicable if the integrals do not converge,
which is the case for a near-threshold S-wave resonance.
An alternative theoretical definition of the resonance energy $E_X$ 
is the center of the resonance in a specified decay mode $k$,
which can be defined mathematically by the condition that its contribution  to the production rate
receives equal contributions from the energy regions  $E< E_X$ and $E> E_X$: 
%===========
\begin{equation}
\int_{E_\mathrm{min}}^{E_X} \!\!\!dE\, \mathrm{Im} [f(E)]^{(k)} =
\int_{E_X}^{E_\mathrm{max}} \!\!\!dE\, \mathrm{Im} [f(E)]^{(k)}.
\label{Eres-def}
\end{equation}
%===========
Our simple prescription  for the resonance is the energy range 
between the specified energies $E_\mathrm{min}$ below the scattering threshold 
and $E_\mathrm{max}$ above the threshold.
An alternative prescription would be to weight the integrals in Eq.~\eqref{Eres-def}
by a Gaussian function of $E$ and extend the limits of the integrals to $-\infty$ and $+\infty$.

In the case of the $X$ resonance, the decay mode that is most convenient 
for defining the resonance energy $E_X$ experimentally  is $J/\psi\, \pi^+\pi^-$. 
This is the decay mode that has been used to obtain the measured value in Eq.~\eqref{EX-exp}.
In the simplest model for the $X$ line shapes defined by the scattering amplitude $f(E)$  in Eq.~\eqref{Amp-E},
the line shape in $J/\psi\, \pi^+\pi^-$ is predicted to be the same as for all the short-distance
decay modes. 
The definition of the resonance energy in Eq.~\eqref{Eres-def} then reduces to
%===========
\begin{equation}
\int_{E_\mathrm{min}}^{E_X} \!dE\, \big|f(E) \big|^2 =
\int_{E_X}^{E_\mathrm{max}} \!dE\, \big|f(E) \big|^2.
\label{EX-def}
\end{equation}
%===========
The resonance factor in Eq.~\eqref{ImAmp-E} at the real energy $E$ can be expressed as
%===========
\begin{eqnarray}
\big|f(E) \big|^2 &= &
\left[ \left( - \mathrm{Re}[\gamma_X]  + \big[\mu (E^2+  \Gamma_{*0}^2/4)^{1/2} - \mu E \big]^{1/2} \right)^2 \right.
\nonumber\\
&& \hspace{1cm}\left.
+ \left( \mathrm{Im}[\gamma_X]  +  \big[ \mu (E^2+  \Gamma_{*0}^2/4)^{1/2} + \mu E \big]^{1/2} \right)^2 \right]^{-1}.
\label{|Amp|^2}
\end{eqnarray}
%===========
All the square roots are of manifestly positive quantities.
At large $|E|$, this resonance factor  decreases as $1/|E|$.
Thus the integrals in the prescription for the resonance energy  $E_X$ in Eq.~\eqref{EX-def} depend logarithmically
on $E_\mathrm{min}$ and $E_\mathrm{max}$.
Our prescription for the resonance as the energy range $E_\mathrm{min} < E < E_\mathrm{max}$
allows us to obtain analytic approximations for the resonance energy 
defined by Eq.~\eqref{EX-def} for three limiting values of $\mathrm{Re}[\gamma_X]$
considered below.

\subsection{Bound-state limit}
\label{sec:Eres-bound}

%%%%%%%%%%%%%%%%%%%%%%%%%%%%%%%%%%%%%%%%%%%%%%%%
\begin{figure}[t]
\includegraphics*[width=0.75\linewidth]{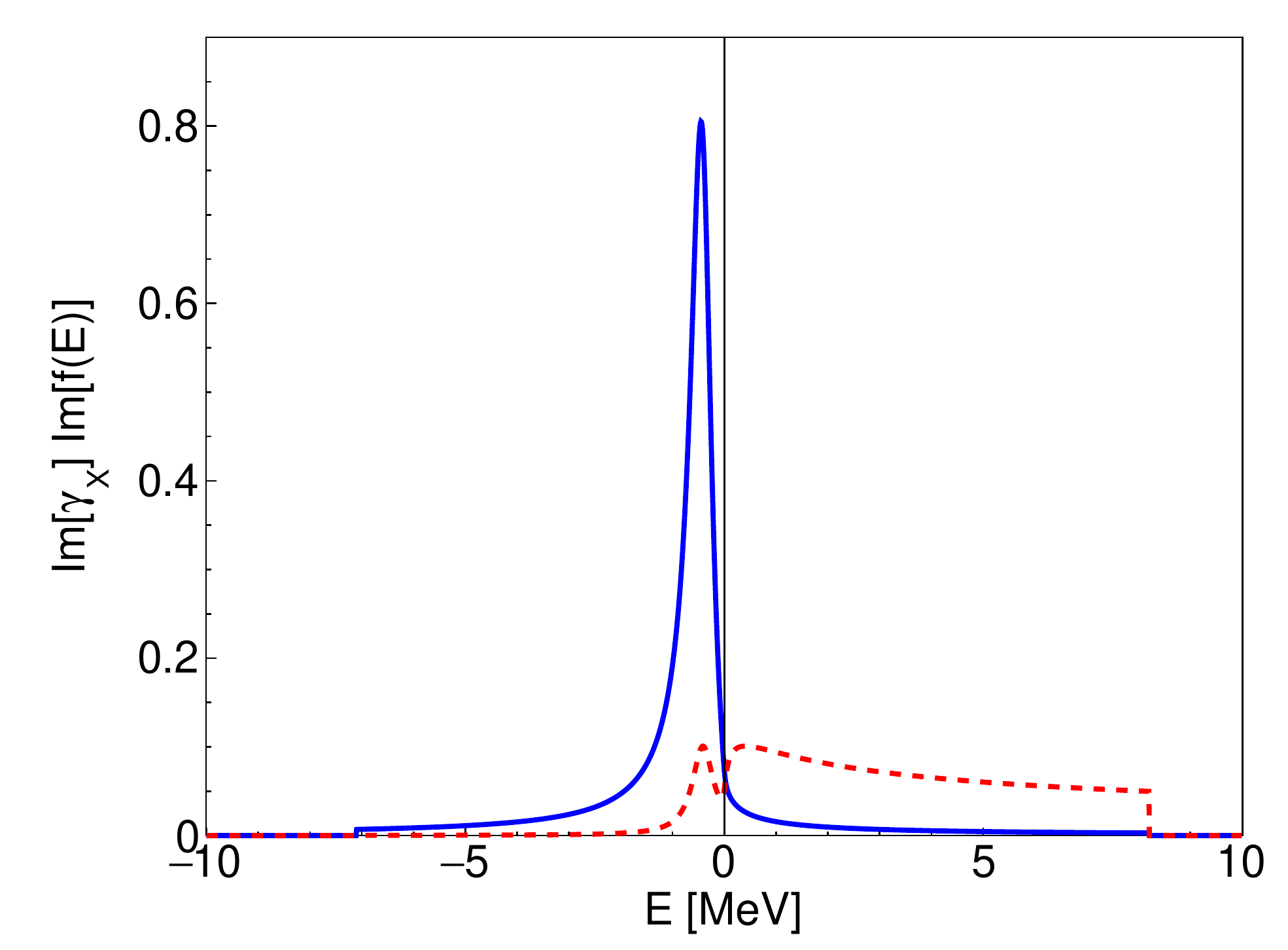} 
\caption{
Line shapes in the SDD channel (solid blue curve) and in the CD channel (dashed red curve) for a bound state.
The parameters are $\mathrm{Re}[\gamma_X]= +4\,  \mathrm{Im}[\gamma_X]$
and $\mathrm{Im}[\gamma_X] = \sqrt{\mu \Gamma_{*0}}$.
The branching fraction of the resonance feature  into the SDD channel inferred from Eq.~\eqref{BR-Xres} is 0.52. 
The branching fraction of the bound state into the SDD channel inferred from Eq.~\eqref{BR-Xbound} is 
8/9.
}
\label{fig:lshbound}
\end{figure}
%%%%%%%%%%%%%%%%%%%%%%%%%%%%%%%%%%%%%%%%%%%%%%

In the  limit $\mathrm{Re}[\gamma_X]^2 \gg \mu \Gamma_{*0}$ with $\mathrm{Re}[\gamma_X] >0$, 
the positive sign of $\mathrm{Re}[\gamma_X]$ implies that the resonance is a bound state.  
The pole energy in Eq.~\eqref{Epole} indicates that its binding energy is approximately $\mathrm{Re}[\gamma_X]^2/2\mu$.
The position of a bound-state pole could be at the  incoming arrow  in  Fig.~\ref{fig:Epolepath},
which is on the physical sheet of  the complex energy plane.
The line shapes in the SDD and CD channels for a bound-state resonance are illustrated in Fig.~\ref{fig:lshbound}.

The SDD line shape is $\mathrm{Im}[\gamma_X]\, |f(E) |^2$,
where the resonance factor is Eq.~\eqref{|Amp|^2}.  
Its properties can be expanded in powers of $1/\mathrm{Re}[\gamma_X]$.
The maximum value of the line shape is $1/\mathrm{Im}[\gamma_X]$,
up to a correction of order $1/\mathrm{Re}[\gamma_X] $.
The maximum is at the energy
%===========
\begin{equation}
 E_\mathrm{SDD} = -\frac{\mathrm{Re}[\gamma_X]^2}{2\mu}
  - \frac{\mathrm{Im}[\gamma_X]\, \Gamma_{*0}}{2\, \mathrm{Re}[\gamma_X]} + \ldots.
\label{Emax-bound} 
\end{equation}
%===========
 The energies where the line shape decreases to half the maximum value are lower by $\Gamma_-$
 and higher by $\Gamma_+$, where 
%===========
\begin{subequations}
\begin{eqnarray}
\Gamma_- &= & \frac12 \Gamma_{*0}
+\frac{2\, \mathrm{Re}[\gamma_X] \, \mathrm{Im}[\gamma_X] + \mathrm{Im}[\gamma_X]^2}{2\mu},
\label{Gamma-}
\\
\Gamma_+ &=&\frac12\Gamma_{*0}
+\frac{ 2\,\mathrm{Re}[\gamma_X] \, \mathrm{Im}[\gamma_X] - \mathrm{Im}[\gamma_X]^2}{2\mu},
\label{Gamma+}
\end{eqnarray}
\label{Gamma+-}%
\end{subequations}
%===========
up to corrections of order $1/\mathrm{Re}[\gamma_X] $.
The full width  in the energy $E$ at half maximum is
%===========
\begin{equation}
\Gamma_X= \Gamma_{*0}+\frac{2\, \mathrm{Re}[\gamma_X] \, \mathrm{Im}[\gamma_X]}{\mu},
\label{GammaX} 
\end{equation}
%===========
up to a correction  of order $1/\mathrm{Re}[\gamma_X]^5$.
This width $\Gamma_X$ coincides with the imaginary part of $-2E_\mathrm{pole}$, 
where the pole energy is given in Eq.~\eqref{Epole}.
 It can be interpreted as the decay rate of the bound state.
The first term on the right side of Eq.~\eqref{GammaX} is the decay rate of the constituent $D^{*0}$ or $\bar D^{*0}$.
The second term can be interpreted as the sum of the partial  decay rates into short-distance decay modes.
Given our assumption that $\mathrm{Im}[\gamma_X]$ is order $\sqrt{\mu \Gamma_{*0}}$,
$\Gamma_X$  is much larger than $\Gamma_{*0}$ in the bound-state limit.

The CD line shape is $|f(E) |^2$ multiplied by the function of $E$ raised to the power $\tfrac12$ in Eq.~\eqref{ImAmp-E}.  
This line shape has two local maxima: 
a bound-state peak below the threshold and a resonance enhancement above the threshold.
The maximum of the bound-state peak in the CD channel  is at the energy 
%===========
\begin{equation}
 E_\mathrm{CD} \approx -\frac{\mathrm{Re}[\gamma_X]^2 - \mathrm{Im}[\gamma_X]^2}{2\mu},
\label{Emax-bound:CD} 
\end{equation}
%===========
 up to a correction of order $1/\mathrm{Re}[\gamma_X]^2$.
 The maximum value of the peak differs from that in the SDD channel 
by a factor of $\mu \Gamma_{*0}/(2\, \mathrm{Re}[\gamma_X] \, \mathrm{Im}[\gamma_X])$, up to a 
relative correction of order  $1/\mathrm{Re}[\gamma_X]^2$. 
The energies where the line shape decreases to half the maximum value are lower by $\Gamma_X/2$
 and higher by $\Gamma_X/2$, up to corrections of order $1/\mathrm{Re}[\gamma_X]$.
Thus the full width  in the energy $E$ at half maximum is essentially the same as that
  for the SDD line shape in Eq.~\eqref{GammaX}.
The ratio of the integrals over $E$ of the bound-state peaks in the SDD and CD channels is therefore 
 roughly equal to the ratio of their maximum values,
which coincides with the ratio of the second and first terms in the expression for the width 
$\Gamma_X$ in Eq.~\eqref{GammaX}.
The SDD over CD branching ratio for the bound state is therefore large in the bound-state limit.

The second peak in the CD line shape comes from the threshold enhancement. 
Its maximum is at an energy near $+\mathrm{Re}[\gamma_X]^2/2\mu$
and its full width in $E$ at half maximum is approximately $4\sqrt{3}\,\mathrm{Re}[\gamma_X]^2/\mu$. 
The ratio of the heights of the peaks from the threshold enhancement and the bound state 
 is approximately $\mathrm{Im}[\gamma_X]^2/\mu \Gamma_{*0}$. 
In Fig.~\ref{fig:Epolepath}, the choice  $\mathrm{Im}[\gamma_X] = \sqrt{\mu \Gamma_{*0}}$ 
ensures that the two peaks have approximately the same height.

The resonance factor in Eq.~\eqref{|Amp|^2} can be approximated by a Breit-Wigner function of
$\sqrt{|E|}$ below the threshold and by a Lorentzian function of $\sqrt{E}$  above the threshold:
%===========
\begin{subequations}
\begin{eqnarray}
\big| f(E) \big|^2 &\approx&
\frac{1}{ \big(\sqrt{2\mu|E|} - \mathrm{Re}[\gamma_X] \big)^2 
+ \big( \mu \Gamma_X/2\,\mathrm{Re}[\gamma_X] \big)^2} 
\qquad E<0,
\label{|Amp|^2-bound<0}%
\\
&\approx&
\frac{1} {\mathrm{Re}[\gamma_X]^2    + \big(\sqrt{2\mu E} + \mu \Gamma_X/2\,\mathrm{Re}[\gamma_X]  \big)^2}
\qquad~~~~~\, E>0.
\label{|Amp|^2-bound>0}
\end{eqnarray}
\label{|Amp|^2-bound}%
\end{subequations}
%===========
The denominator of the Breit-Wigner function in Eq.~\eqref{|Amp|^2-bound<0} has errors 
of order $1/ \mathrm{Re}[\gamma_X]$ for negative $E$ of order $\mathrm{Re}[\gamma_X]^2/\mu$.
It implies that the bound state has binding energy $\mathrm{Re}[\gamma_X]^2/2\mu$ 
and that its full width in $E$ at half maximum is $\Gamma_X$ in Eq.~\eqref{GammaX}.
The denominator of the Lorentzian function in Eq.~\eqref{|Amp|^2-bound>0} has errors that are 
zeroth order in $ \mathrm{Re}[\gamma_X]$ for positive $E$ of order $\mathrm{Re}[\gamma_X]^2/\mu$.
The pole approximation in Eq.~\eqref{f-E:pole} gives a good approximation 
to the resonance factor $|f(E)|^2$ at real energies $E$ near the bound-state peak.  
It predicts the maximum and the energy at the maximum with relative errors of order $1/\mathrm{Re}[\gamma_X]^2$,
and it predicts the full width at half maximum with a relative error of order $1/\mathrm{Re}[\gamma_X]^5$.

The resonance energy defined by Eq.~\eqref{EX-def}  can be calculated using the approximation 
for  $|f(E)|^2$ in Eq.~\eqref{|Amp|^2-bound} and then expanded in powers of $\Gamma_X$.
The resonance energy is
%===========
\begin{equation}
 E_X \approx -\frac{\mathrm{Re}[\gamma_X]^2}{2\mu}+ 
\left[ \log\frac{E_\mathrm{max}}{|E_\mathrm{min}|} - 4\log\frac{2\, \mathrm{Re}[\gamma_X]^2}{\mu \Gamma_X} - 2 \right] 
\frac{\mu \Gamma_X^2}{16\, \mathrm{Re}[\gamma_X]^2} +\ldots.
\label{EX-bound} 
\end{equation}
%===========
We have simplified the coefficient of the leading correction term by taking the limit 
$|E_\mathrm{min}| \gg \mathrm{Re}[\gamma_X]^2/\mu$.
The coefficient depends logarithmically on $E_\mathrm{max}/|E_\mathrm{min}|$
and on $\mathrm{Re}[\gamma_X]^2/\mu \Gamma_X$.

\subsection{Zero-energy resonance}

%%%%%%%%%%%%%%%%%%%%%%%%%%%%%%%%%%%%%%%%%%%%%%%%
\begin{figure}[t]
\includegraphics*[width=0.75\linewidth]{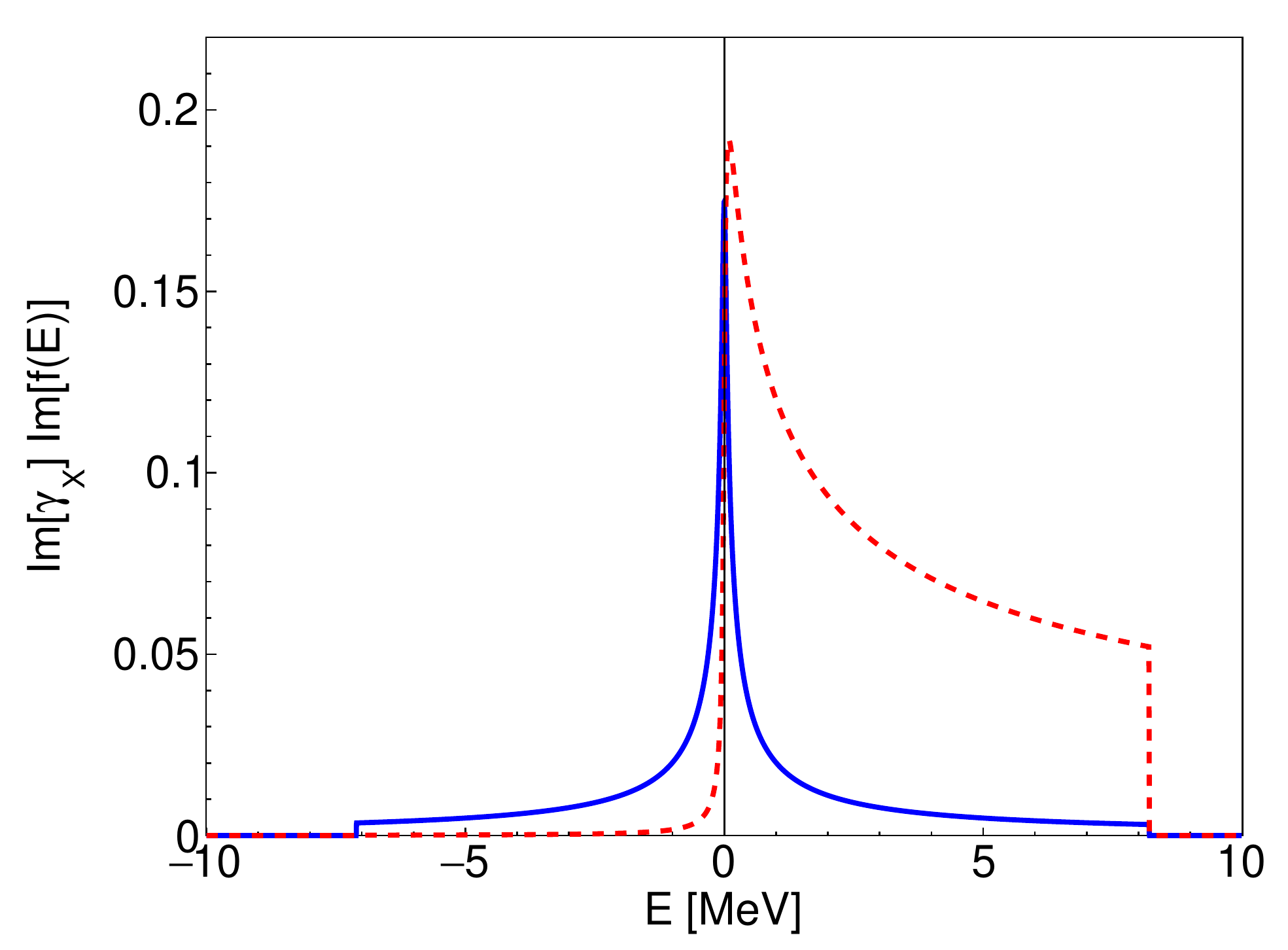} 
\caption{
Line shapes in the  SDD channel (solid blue curve) and in the CD channel (dashed red curve) 
for a zero-energy resonance.  The parameters are
$\mathrm{Re}[\gamma_X]= -\mathrm{Im}[\gamma_X]$ and $\mathrm{Im}[\gamma_X] = \sqrt{\mu \Gamma_{*0}}$.
The branching fraction of the resonance feature  into the SDD channel inferred from Eq.~\eqref{BR-Xres}
 is  0.22. 
}
\label{fig:lsh0}
\end{figure}
%%%%%%%%%%%%%%%%%%%%%%%%%%%%%%%%%%%%%%%%%%%%%%

In the case $\mathrm{Re}[\gamma_X] = -\mathrm{Im}[\gamma_X]$, 
 the negative sign of $\mathrm{Re}[\gamma_X]$ implies that the resonance is a virtual state.
The pole energy in Eq.~\eqref{Epole} indicates that the complex energy of the virtual state is order $\Gamma_{*0}$.
The pole is on the second sheet of the complex energy plane, as illustrated in  Fig.~\ref{fig:Epolepath}.
The line shapes in the SDD and CD channels 
for a zero-energy resonance are illustrated in Fig.~\ref{fig:lsh0}.

The resonance factor in Eq.~\eqref{|Amp|^2} is an even function of $E$ whose maximum is at $E = 0$. 
The maximum value is 
 %===========
\begin{equation}
\big| f(0) \big|^2 =
\frac{2}{\big[ 2\, \mathrm{Im}[\gamma_X] + \sqrt{2\mu\Gamma_{*0}} \,\big]^2} .
\label{fsq-zero} 
\end{equation}
%===========
The full width in $E$ at half maximum is order $\Gamma_{*0}$.
It changes smoothly from $1.73~\Gamma_{*0}$ if $\mathrm{Im}[\gamma_X] \ll \sqrt{\mu\Gamma_{*0}}$ 
to $3.75~\Gamma_{*0}$ if $\mathrm{Im}[\gamma_X]= \sqrt{\mu\Gamma_{*0}/2}$ 
to $0.54~\mathrm{Im}[\gamma_X]^2/\mu$ if $\mathrm{Im}[\gamma_X] \gg \sqrt{\mu\Gamma_{*0}}$.
The pole approximation in Eq.~\eqref{f-E:pole} predicts correctly that the maximum of 
$|f(E)|^2$ is at  $E=0$, but it does not a good  approximation to the shape of the peak.  
It predicts, for example, that  for $\mathrm{Im}[\gamma_X] =  \sqrt{\mu\Gamma_{*0}/2}$,
the maximum value is $\infty$ and the full width at half maximum is 0.

We can deduce an expression for the resonance energy $E_X$ by exploiting the fact that 
$f(E)$ is an even function of $E$, which implies that integrals of $|f(E)|^2$ satisfy
%===========
\begin{equation}
\int_{-E_0}^{0} \!dE\, \big|f(E) \big|^2 =
\int_{0}^{E_0} \!dE\, \big|f(E) \big|^2
\label{int-E0}
\end{equation}
%===========
for any $E_0$.  By assuming $|E_X| \ll \Gamma_{*0}$ and using Eq.~\eqref{int-E0} with $E_0=|E_\mathrm{min}|$,
we can reduce Eq.~\eqref{EX-def} for $E_X$ to an integral over the small-$E$ region where $|f(E)|^2$
reduces to $|f(0)|^2$ and an integral over the large-$E$ region where $|f(E)|^2$ reduces to $1/(2|E|)$.
The resulting approximation for the resonance energy is
%===========
\begin{equation}
E_X  \approx \left(\log\frac{E_\mathrm{max}}{|E_\mathrm{min}|}  \right)
\frac{\big( 2\, \mathrm{Im}[\gamma_X] + \sqrt{2\mu \Gamma_{*0}} \,\big)^2}{8\mu}.
\label{EXzero}
\end{equation}
%===========
This result is valid provided the logarithm is small compared to 1, so that $|E_X| \ll \Gamma_{*0}$. 
This expression for $E_X$ can be positive or negative 
depending on whether $E_\mathrm{max}$ is larger or smaller than $|E_\mathrm{min}|$.

\subsection{Virtual-state limit}

%%%%%%%%%%%%%%%%%%%%%%%%%%%%%%%%%%%%%%%%%%%%%%%%
\begin{figure}[t]
\includegraphics*[width=0.75\linewidth]{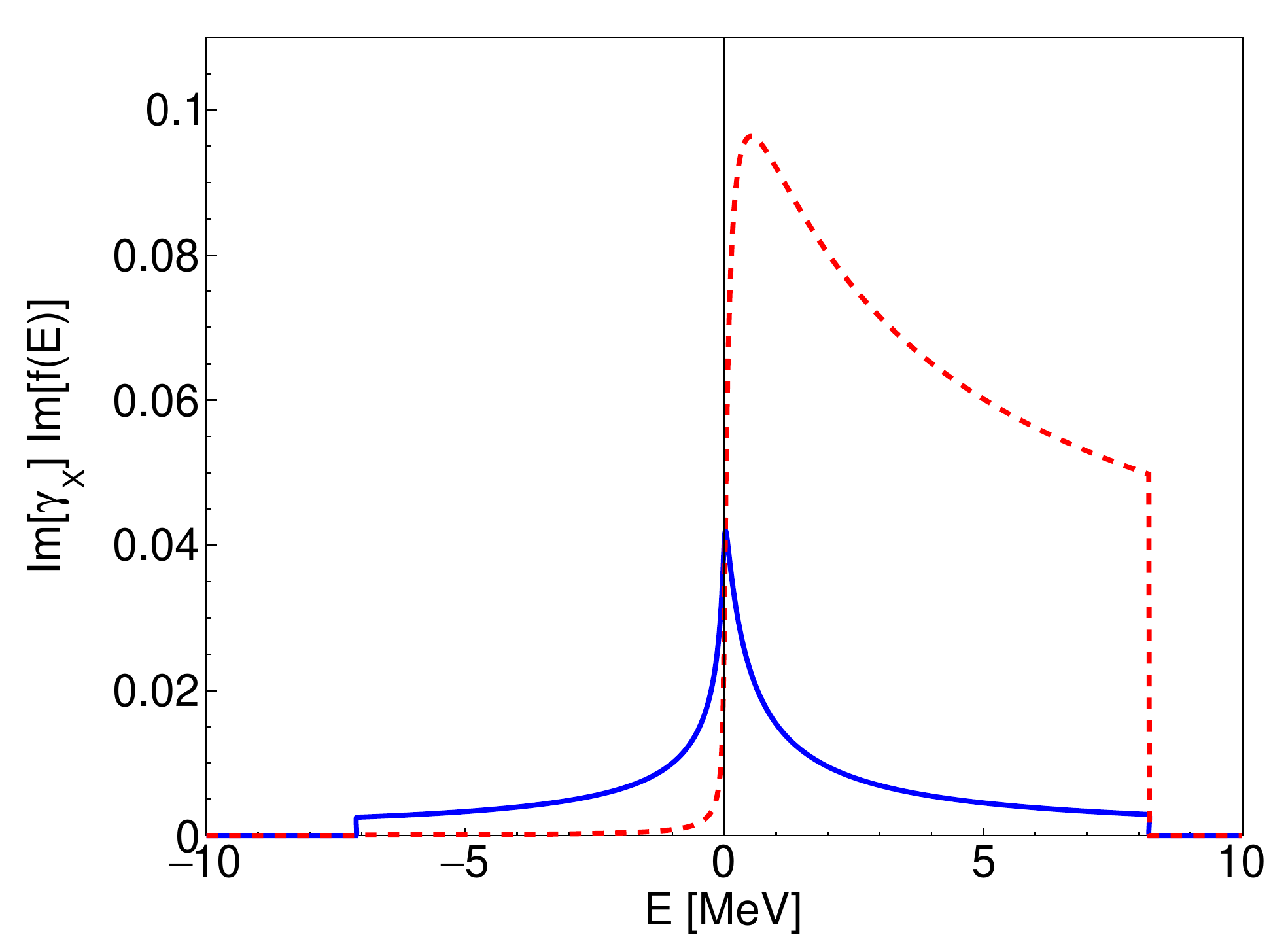} 
\caption{
Line shapes in the SDD channel (solid blue curve) and in the CD channel (dashed red curve) for a virtual state. 
The parameters are $\mathrm{Re}[\gamma_X]= -4\,  \mathrm{Im}[\gamma_X]$
and $\mathrm{Im}[\gamma_X] = \sqrt{\mu \Gamma_{*0}}$.
The branching fraction of the resonance feature  into the SDD channel inferred from Eq.~\eqref{BR-Xres}
is  0.17. 
}
\label{fig:lshvirtual}
\end{figure}
%%%%%%%%%%%%%%%%%%%%%%%%%%%%%%%%%%%%%%%%%%%%%%

In the  limit $\mathrm{Re}[\gamma_X]^2 \gg \mu \Gamma_{*0}$ with $\mathrm{Re}[\gamma_X] <0$, 
the negative sign of $\mathrm{Re}[\gamma_X]$ implies that the resonance is associated with a virtual state.  
The pole energy in Eq.~\eqref{Epole} indicates that the virtual state  has a negative energy 
that is approximately $-\mathrm{Re}[\gamma_X]^2/2\mu$.
However the resonance energy $E_X$, which is also of order $\mathrm{Re}[\gamma_X]^2/\mu$, is positive.
The position of a virtual-state pole could be at the  outgoing arrow  in  Fig.~\ref{fig:Epolepath},
which is on the second sheet of  the complex energy plane.
The line shapes in the SDD and CD channels 
for a virtual-state resonance are illustrated in Fig.~\ref{fig:lshvirtual}.

The resonance factor in Eq.~\eqref{|Amp|^2} can be simplified by setting 
$\Gamma_{*0} = 0$ and $\mathrm{Im}[\gamma_X] = 0$:
%===========
\begin{subequations}
\begin{eqnarray}
\big| f(E) \big|^2 &\approx&
\frac{1}{ \big(\sqrt{2\mu|E|} + \big| \mathrm{Re}[\gamma_X] \big| \big)^2} 
\qquad E<0,
\\ 
 &\approx& \frac{1}{\mathrm{Re}[\gamma_X]^2 + 2 \mu E} 
  \qquad   \qquad ~~~~E >0.
\end{eqnarray}
\label{|Amp|^2-virtual}%
\end{subequations}
%===========
Its maximum is  at $E=0$.   
It decreases to half the maximum at the energies $-0.086~\mathrm{Re}[\gamma_X]^2/\mu$
and $+\mathrm{Re}[\gamma_X]^2/2\mu$,  so
its full width at half maximum is $0.586~\mathrm{Re}[\gamma_X]^2/\mu$. 
The pole approximation in Eq.~\eqref{f-E:pole} gives a very bad approximation 
to the resonance factor $|f(E)|^2$ at real energies $E$.  
It predicts a maximum at an energy near $-\mathrm{Re}[\gamma_X]^2/2\mu$ instead of at 0.

The resonance energy defined by Eq.~\eqref{EX-def} can be 
obtained by evaluating integrals of the resonance factor in Eq.~\eqref{|Amp|^2-virtual}.  
Assuming $|E_\mathrm{min}|, E_\mathrm{max} \gg \mathrm{Re}[\gamma_X]^2/\mu$,
the resonance energy can be approximated by
%===========
\begin{equation}
E_X  \approx \left(e \, \sqrt{\frac{E_\mathrm{max}}{|E_\mathrm{min}|}}  - 1 \right)
\frac{ \mathrm{Re}[\gamma_X]^2}{2\mu}.
\label{EXvirtual}
\end{equation}
%===========
This expression for  $E_X$ is positive provided $E_\mathrm{max} > 0.141\, |E_\mathrm{min}|$.

%\newpage 

%%%%%%%%%%%%%%%%%%%%%%%%%%%%%%%%%%%%
\section{Branching fractions}
\label{sec:Branching}
%%%%%%%%%%%%%%%%%%%%%%%%%%%%%%%%%%%%

The difference between the line shapes of a near-threshold S-wave resonance 
and that of  a conventional resonance complicates the definition of the $X$ branching fractions.  
In this Section, we emphasize the difference between branching fractions for an 
$X$ resonance feature and branching fractions for the $X$ bound state.

\subsection{Branching fractions for the $\bm{X}$ resonance feature}

The BaBar collaboration has recently determined the inclusive branching fraction of the $B^+$ meson  into 
$K^+$ plus the $X$ resonance by measuring  the recoil momentum spectrum of the $K^+$ \cite{Wormser}.
There is a peak in the momentum spectrum that corresponds to the recoil of the $K^+$
against a system with invariant mass near 3872~MeV.
The recoiling system includes the threshold enhancement in $D^{*0} \bar D^0$ and $D^0 \bar D^{*0}$ above the threshold 
as well as the decay products of a possible $X$ bound state below the threshold  or a virtual state.
We refer to the entire recoiling system as the {\it $X$ resonance feature}
from $B^+ \to K^+$ transitions, and we denote it by $X_{B^+ \to K^+}$.
The preliminary result for the inclusive branching fraction into the $X$ resonance feature,
with errors combined in quadrature, is \cite{Wormser}
%===========
\begin{equation}
\mathrm{Br}[B^+ \to K^+ X_{B^+ \to K^+}] = (2.1 \pm 0.7)\times 10^{-4}.
\label{BrBKXres}
\end{equation}
%===========
This result is consistent with previous upper bounds by the  BaBar collaboration  \cite{Aubert:2005vi}
and by the Belle collaboration \cite{Kato:2017gfv}.
If the measured product branching fraction for $B^+ \to K^+ (J/\psi \, \pi^+\pi^-)$ \cite{Tanabashi:2018oca} 
is divided by the inclusive branching fraction in Eq.~\eqref{BrBKXres}, it gives   \cite{Wormser}
%===========
\begin{equation}
\mathrm{Br}[X_{B^+ \to K^+} \to J/\psi \, \pi^+\pi^-]  \equiv 
\frac{\mathrm{Br}[B^+ \to K^+ X] \, \mathrm{Br}[X \to J/\psi \, \pi^+\pi^-]}
       {\mathrm{Br}[B^+ \to K^+ X_{B^+ \to K^+}]}
= (4.1 \pm 1.3)\%.
\label{BrXresJpsipipi}
\end{equation}
%===========
We refer to this as the $J/\psi \, \pi^+\pi^-$ branching fraction of the $X$ resonance feature 
from $B^+$-to-$K^+$ transitions.
This branching fraction need not be the same for other production mechanisms, 
such as $B^0$-to-$K^0$ transitions or $e^+e^-$-to-$\gamma$ transitions.

If the resonance is associated with  a bound state,
the branching fraction of the $X$ resonance feature in Eq.~\eqref{BrXresJpsipipi}
should be distinguished from the $J/\psi \, \pi^+\pi^-$ branching fraction of the $X$ bound state,
which we denote by $\mathrm{Br}[X \to J/\psi \, \pi^+\pi^-] $ with no subscript on $X$.
This larger branching fraction could be obtained by dividing 
 the product branching fraction for $B^+ \to K^+ (J/\psi \, \pi^+\pi^-)$
by the branching fraction of $B^+$ into $K^+$ plus the $X$  bound state, 
which we denote by $\mathrm{Br}[B^+ \to K^+ X]$ with no subscript on $X$.
This branching fraction into the $X$ bound state has not been measured.
The branching fractions of the $X$ bound state should be the same for all short-distance production mechanisms, 
because they can be obtained by factoring the production amplitudes at the bound-state pole
as in Eq.~\eqref{dR/dE:factor}.

The  $D^0 \bar D^0 \pi^0$ branching fraction of the $X$ resonance feature from $B^+$-to-$K^+$ transitions 
can be obtained using measurements by the Belle collaboration of the 
decay $B^+ \to K^+ (D^0 \bar D^0 \pi^0)$ \cite{Gokhroo:2006bt}.
The $D^0 \bar D^0 \pi^0$ invariant mass distribution has a narrow peak near the $D^{*0} \bar D^0$ threshold
that can be identified with the $X$ resonance feature.
The measured branching fraction for events in the peak can  be interpreted as the contribution 
to the inclusive branching fraction into the $X$ resonance feature in Eq.~\eqref{BrBKXres}
from the final state $D^0 \bar D^0 \pi^0$.
Dividing  by that inclusive branching fraction, we obtain the $D^0 \bar D^0 \pi^0$ branching fraction 
of the $X$ resonance feature from $B^+$-to-$K^+$ transitions:
%===========
\begin{equation}
\mathrm{Br}[X_{B^+ \to K^+}  \to D^0 \bar D^0 \pi^0]
 \equiv \frac{\mathrm{Br}[B^+ \to K^+(D^0 \bar D^0 \pi^0)]}
{\mathrm{Br}[B^+ \to K^+ X_{B^+ \to K^+}] } = (49\pm 26)\%,
\label{BRDD*Belle}
\end{equation}
%===========
where we have combined the errors in quadrature.
The CD branching  fraction is the sum of the branching fractions into
$D^0 \bar D^0 \pi^0$ and  $D^0 \bar D^0 \gamma$.
The additional contribution from the final state $D^0 \bar D^0 \gamma$ can be taken into account
approximately by dividing the branching fraction in Eq.~\eqref{BRDD*Belle}
by the known branching fraction of $D^{*0}$ into $D^0 \pi^0$. 
The resulting estimate for the CD branching fraction of the $X$ resonance feature from $B^+$-to-$K^+$ transitions is  
%===========
\begin{equation}
\mathrm{Br}[X_{B^+ \to K^+}  \to D^0 \bar D^0 (\pi^0,\gamma)]
 \approx \frac{\mathrm{Br}[X_{B^+ \to K^+}  \to D^0 \bar D^0 \pi^0]}
{\mathrm{Br}[D^{*0} \to D^0\pi^0] } = (75\pm 40)\%.
\label{BrDDpigamma}
\end{equation}
%===========
The corresponding estimate for the SDD branching fraction is  the complimentary fraction $(25\pm 40)\%$.

Assuming decays of the $X$  resonance feature into $J/\psi \, \pi^+\pi^-$ are dominated
by decays of the $X$ bound state, we can obtain an estimate of the $D^0 \bar D^0 (\pi^0,\gamma)$ over $J/\psi\, \pi^+\pi^-$ 
branching ratio for the $X$ resonance feature from $B^+$-to-$K^+$ transitions by dividing
the Belle result for the  product branching fraction
for $B^+ \to K^+ (D^0 \bar D^0 \pi^0)$  \cite{Gokhroo:2006bt} by the measured product branching fraction
for $B^+ \to K^+ (J/\psi \, \pi^+\pi^-)$ \cite{Tanabashi:2018oca}
and by the branching fraction for $D^{*0} \to D^0 \pi^0$:
%===========
\begin{equation}
\frac{\mathrm{Br}[X_{B^+ \to K^+}  \to D^0 \bar D^0 (\pi^0,\gamma)]}
{\mathrm{Br}[X_{B^+ \to K^+}  \to J/\psi\, \pi^+\pi^-]}
  = 18.3 \pm 7.8.
\label{BRDDpigamma/psipipi}
\end{equation}
%===========

Estimates of the CD branching fraction of the $X$ resonance feature from $B^+$-to-$K^+$ transitions
can also be obtained from measurements by the BaBar and Belle collaborations 
of the decays of $B^+$ into $K^+$ plus $D^{*0} \bar D^0$ \cite{Aubert:2007rva,Adachi:2008sua}.
These measurements are complicated by systematic errors associated with constraining the momenta
of $D^0$ and $\pi^0$ with invariant mass close to the mass of the $D^{*0}$ so their invariant mass 
is exactly $M_{*0}$ \cite{Stapleton:2009ey}.  This procedure moves $D^0 \bar D^0 \pi^0$ events 
with invariant mass below the $D^{*0} \bar D^0$ threshold to above the threshold.
The resulting measurement of the resonance energy of the $X$ in the $D^0 \bar D^{*0}$ decay mode
therefore has an  undetermined positive systematic error  \cite{Stapleton:2009ey}.
The complications from constraining  momenta so the invariant mass of $D^0\pi^0$ or $\bar D^0\pi^0$ is $M_{*0}$
affects a measurement of the energy and width of the resonance 
more than a measurement of the branching fraction.
We can therefore  interpret  a measured product branching fraction for the decay 
$B^+ \to K^+ \, (D^{*0} \bar D^0,D^0 \bar D^{*0})$ as an approximation to that for the decay  
$B^+ \to K^+ \, D^0 \bar D^0 (\pi^0,\gamma)$.
A branching fraction for the $X$ resonance feature into $D^0 \bar D^0 (\pi^0,\gamma)$ can then be obtained 
by dividing the measured product branching fraction  by the inclusive branching fraction in Eq.~\eqref{BrBKXres}.
The BaBar measurement in Ref.~\cite{Aubert:2007rva} gives the estimate $(80\pm 38)\%$, 
which is consistent with the branching fraction in Eq.~\eqref{BrDDpigamma}.
The Belle  measurement in Ref.~\cite{Adachi:2008sua} gives the estimate $(37\pm 15)\%$, 
whose central value is significantly smaller than that in  Eq.~\eqref{BrDDpigamma}.

\subsection{Branching fractions for the $\bm{X}$ bound state}

If $X$ is a narrow bound state whose width is sufficiently small compared to its binding energy,
it has well-defined branching fractions into its various decay modes
that do not depend on the production mechanism.
The branching fractions into short-distance decay modes
should be considerably larger than the corresponding branching fractions for an $X$ resonance feature.
The $J/\psi \, \pi^+\pi^-$ branching fraction of the $X$ resonance feature from $B^+$-to-$K^+$ transitions  in Eq.~\eqref{BrXresJpsipipi}
can be taken as a loose lower bound on the $J/\psi\, \pi^+\pi^-$ branching fraction for the $X$ bound state.
An upper bound on the $J/\psi\, \pi^+\pi^-$ branching fraction for the $X$ bound state  
can be obtained from  measurements of branching ratios for other decay modes.
The bound is obtained most easily by using the equality
%===============
\begin{equation}
\frac{1}{\mathrm{Br} [X \to J/\psi\, \pi^+\pi^-]} =
1 + {\sum_i}' \frac{\mathrm{Br} [X \to i]}{\mathrm{Br} [X \to J/\psi\, \pi^+\pi^-]} 
+ \frac{\mathrm{Br} [X \to D^0 \bar D^0 (\pi^0,\gamma)]}{\mathrm{Br} [X \to J/\psi\, \pi^+\pi^-]} ,
\label{1/BrJpsipipi}
\end{equation}
%===============
where the sum over $i$ is over SDD modes other than $J/\psi\, \pi^+\pi^-$.
The other SDD modes that have been observed 
are $J/\psi\, \pi^+\pi^-\pi^0$, $J/\psi\, \gamma$, $\psi(2S)\, \gamma$,  and $\chi_{c1}\, \pi^0$.
The branching ratio on the right side of Eq.~\eqref{1/BrJpsipipi} for  
$i=J/\psi  \,\pi^+\pi^-\pi^0$ (which is often called $J/\psi \,\omega$)
is $0.8 \pm 0.3$ \cite{delAmoSanchez:2010jr}.
The branching ratio  for $i=J/\psi \,\gamma$ is determined   to be $0.24\pm 0.05$ by calculating the ratio of product
 branching fractions from $B^+ \to K^+X$ decays \cite{Tanabashi:2018oca}.
The branching ratio  for $i=\psi(2S) \,\gamma$ can be obtained from that for $J/\psi \,\gamma$
by multiplying by  the  branching ratio $2.6\pm 0.6$ for the decays of $X$ into 
$\psi(2S) \,\gamma$ over $J/\psi \,\gamma$  \cite{Tanabashi:2018oca}.
The branching ratio  for the recently observed decay mode $i=\chi_{c1} \,\pi^0$ is $0.88\pm 0.34$  \cite{Ablikim:2019soz}.
With these  four decay modes included but the last term in Eq.~\eqref{1/BrJpsipipi} excluded,
the right side of Eq.~\eqref{1/BrJpsipipi} is $3.56 \pm 0.51$.
 Its reciprocal is $(28.1\pm 4.1)\%$. 
An upper bound on the branching fraction with 90\% confidence level can be obtained 
by adding $1.28\,\sigma$ to the central value: 
%===============
\begin{equation}
\mathrm{Br} [X\to J/\psi\, \pi^+\pi^-] < 33\%  \qquad (90\%~\mathrm{C.L.}).
\label{BrJpsipipi<}
\end{equation}
%===============
This upper bound would be 44\% if the decay mode  $\chi_{c1} \,\pi^0$ was not taken into account.
The $D^0 \bar D^0 (\pi^0,\gamma)$ term in Eq.~\eqref{1/BrJpsipipi} cannot be taken into account,
because there are no measurements of this branching ratio for the $X$ bound state.

The identity in Eq.~\eqref{1/BrJpsipipi} holds equally well if all the branching fractions for the $X$ bound state 
are replaced by branching fractions for an $X$ resonance feature.  
We can use this identity to obtain an upper bound on the $J/\psi\, \pi^+\pi^-$ branching fraction of the $X$ resonance feature 
from $B^+$-to-$K^+$ transitions.  
We assume the branching ratio of the $X$ resonance feature for each SDD mode $i$ over $J/\psi\, \pi^+\pi^-$
is the same as the corresponding branching ratio of the $X$ bound state.
An estimate for the $D^0 \bar D^0 (\pi^0,\gamma)$ over $J/\psi\, \pi^+\pi^-$ branching ratio 
for the $X$ resonance feature from $B^+$-to-$K^+$ transitions is given in Eq.~\eqref{BRDDpigamma/psipipi}.
If  the last term in Eq.~\eqref{1/BrJpsipipi} is replaced by this value, the right side becomes 
$21.9 \pm 7.8$. Its reciprocal is  $(4.6 \pm 1.6)\%$. 
An upper bound with 90\% confidence level on the $J/\psi\, \pi^+\pi^-$ branching fraction of the $X$ resonance feature 
from $B^+$-to-$K^+$ transitions can be obtained by adding $1.28\,\sigma$ to the central value:
%===============
\begin{equation}
\mathrm{Br} [X_{B^+ \to K^+} \to J/\psi\, \pi^+\pi^-] < 6.7\%  \qquad (90\%~\mathrm{C.L.}).
\label{BrJpsipipi}
\end{equation}
%===============
This upper bound is consistent with the branching fraction in Eq.~\eqref{BrXresJpsipipi}
 determined recently by the BaBar collaboration \cite{Wormser}.

Upper bounds on the branching fraction for $X$ into $J/\psi\, \pi^+\pi^-$ 
that are significantly smaller than that in Eq.~\eqref{BrJpsipipi<} have been derived previously,
but they are actually upper bounds on the branching fraction for an $X$ resonance feature into $J/\psi\, \pi^+\pi^-$.
The upper bound without taking into account the decay mode  $\chi_{c1} \,\pi^0$  
was given as 8.3\% in Ref.~\cite{Guo:2014sca} and 10\% in Ref.~\cite{Yuan:2018inv}.
These are much smaller than our upper bound of 44\% without taking into account $\chi_{c1} \,\pi^0$.
A smaller upper bound can be obtained by approximating the last term
in Eq.~\eqref{1/BrJpsipipi} by the ratio $9.9\pm 3.2$ 
of the PDG values for the product branching fractions for $B^+ \to K^+(\bar D^{*0} D^0 )$ 
and  $B^+ \to K^+(J/\psi\, \pi^+\pi^-)$ \cite{Tanabashi:2018oca}.
With this additional contribution to the right side of  Eq.~\eqref{1/BrJpsipipi}
but with the $\chi_{c1} \,\pi^0$ term excluded, it becomes $12.6\pm 3.2$. 
Its reciprocal is $(8.0\pm 2.0)\%$. 
The central value is consistent with the upper bound  in Ref.~\cite{Guo:2014sca}.
Adding $1.28\,\sigma$ to the central value  to get an upper bound at the 90\% confidence level
gives  10.5\%, which is close to the upper bound in Ref.~\cite{Yuan:2018inv}.
Having used a product branching fraction for $B^+ \to K^+(\bar D^{*0} D^0 )$ as an input,
the upper bounds in Refs.~\cite{Guo:2014sca} and \cite{Yuan:2018inv}
are actually on the $J/\psi\, \pi^+\pi^-$ branching fraction of the $X$ resonance feature 
from $B^+$-to-$K^+$ transitions.

\subsection{Theoretical branching fractions}

Because the $X$  line shapes  differ from those of a conventional resonance,
theoretical expressions for the branching fractions of the $X$ resonance require a prescription.  
The branching ratio of a resonance feature into the final state $k$ over the final state $l$
can be expressed as a ratio of integrals of the corresponding contributions to the inclusive line shape,
which has the form in Eq.~\eqref{dR/dE:optical}:
%===========
\begin{equation}
\frac{\mathrm{Br}[X_\mathrm{res}  \to k]}
{\mathrm{Br}[X_\mathrm{res}  \to l]}
 = \frac{ \sum_{ij} B_i {B_j}^* \int_{E_\mathrm{min}}^{E_\mathrm{max}} \!dE\, \mathrm{Im}[f_{ij}(E)]^{(k)}}
{ \sum_{ij} B_i {B_j}^* \int_{E_\mathrm{min}}^{E_\mathrm{max}} \!dE\, \mathrm{Im}[f_{ij}(E)]^{(l)}} .
\label{BR-Xij}
\end{equation}
%===========
We have used the simple prescription  for the resonance as the energy range 
between the specified energies $E_\mathrm{min}$ below the scattering threshold 
and $E_\mathrm{max}$ above the threshold.
An alternative prescription would be to weight the integrals in Eq.~\eqref{BR-Xij}
by a Gaussian function of $E$ and extend the limits of the integrals to $-\infty$ and $+\infty$.
The choice of the resonance feature, which is represented by $X_\mathrm{res}$,
determines the short-distance coefficients $B_i$.
 
In the simplest model for the $X$  line shapes,
there is a single resonant scattering channel with the scattering amplitude $f(E)$  in Eq.~\eqref{Amp-E}.
The inclusive line shape is proportional to the function  $\mathrm{Im}[f(E)]$ in Eq.~\eqref{ImAmp-E}.
The SDD over CD branching ratio of the $X$ resonance feature can be expressed as a ratio of integrals:
%===========
\begin{equation}
\frac{\mathrm{Br}[X_\mathrm{res}  \to \mathrm{SDD}]}
{\mathrm{Br}[X_\mathrm{res}  \to D^0 \bar D^0 (\pi^0,\gamma)]}
 = \frac{ \mathrm{Im}[\gamma_X]\,
 \int_{E_\mathrm{min}}^{E_\mathrm{max}} \!dE\, \big| f(E) \big|^2}
{\int_{E_\mathrm{min}}^{E_\mathrm{max}} \!dE\, \big| f(E) \big|^2
\big[\mu \sqrt{E^2+  \Gamma_{*0}^2/4} + \mu E \big]^{1/2}} .
\label{BR-Xres}
\end{equation}
%===========
The numerator depends logarithmically on $E_\mathrm{min}$ and $E_\mathrm{max}$.
The denominator is insensitive to $E_\mathrm{min}$, but it depends on the upper endpoint as $E_\mathrm{max}^{1/2}$. 
The  integrals in Eq.~\eqref{BR-Xres} are complicated functions  
of $\Gamma_{*0}$ and the real and imaginary parts of $\gamma_X$.
Given an SDD over CD branching ratio BR, such as that in Eq.~\eqref{BR-Xres},
the SDD branching fraction is $\mathrm{BF} = \mathrm{BR}/(1+\mathrm{BR})$.

If the $X$ is  a narrow bound state whose width is sufficiently small compared to its binding energy,
the branching fractions for decays of the bound state are well defined
and they do not depend on the production mechanism.
They give the fractions of events that would be observed if the energy $E$ could be tuned to 
closer to the negative resonance energy $E_X$ than the half-width $\Gamma_X/2$.
In the simplest model for the $X$ line shapes,
the SDD over CD branching ratio for the $X$ bound state can be approximated by
the ratio of the two terms in the expression for $\Gamma_X$ in Eq.~\eqref{GammaX}:
%===========
\begin{equation}
\frac{\mathrm{Br}[X  \to  \mathrm{SDD}]}
{\mathrm{Br}[X  \to D^0 \bar D^0 (\pi^0,\gamma)]}
 \approx \frac{2\,  \mathrm{Re}[\gamma_X]\,  \mathrm{Im}[\gamma_X]}{\mu  \Gamma_{*0}} .
\label{BR-Xbound}
\end{equation}
%===========
This is a simple function  of $\Gamma_{*0}$ and the real and imaginary parts of $\gamma_X$.
This expression for the branching ratio has an unphysical negative value when  $\mathrm{Re}[\gamma_X]$
is negative. It is therefore a reasonable approximation only $\mathrm{Re}[\gamma_X]$
is large enough, corresponding to a sufficiently narrow bound state.
As a reasonable quantitative criterion for the validity of Eq.~\eqref{BR-Xbound},
we choose to require the real part of the pole energy $E_\mathrm{pole}$ in Eq.~\eqref{Epole} 
to be larger in absolute value than its imaginary part. 
This criterion implies that the boundary of the region where the bound state is sufficiently narrow 
is when the real part of $\gamma_X$ is
%===========
\begin{equation}
\mathrm{Re}[\gamma_X] = \mathrm{Im}[\gamma_X]
+ \sqrt{ 2\,  \mathrm{Im}[\gamma_X]^2 + \mu \Gamma_{*0} } .
\label{Regamma:nbs}
\end{equation}
%===========

%%%%%%%%%%%%%%%%%%%%%%%%%%%%%%%%%%%%%%%%%%%%%%%%
\begin{figure}[t]
\includegraphics*[width=0.80\linewidth]{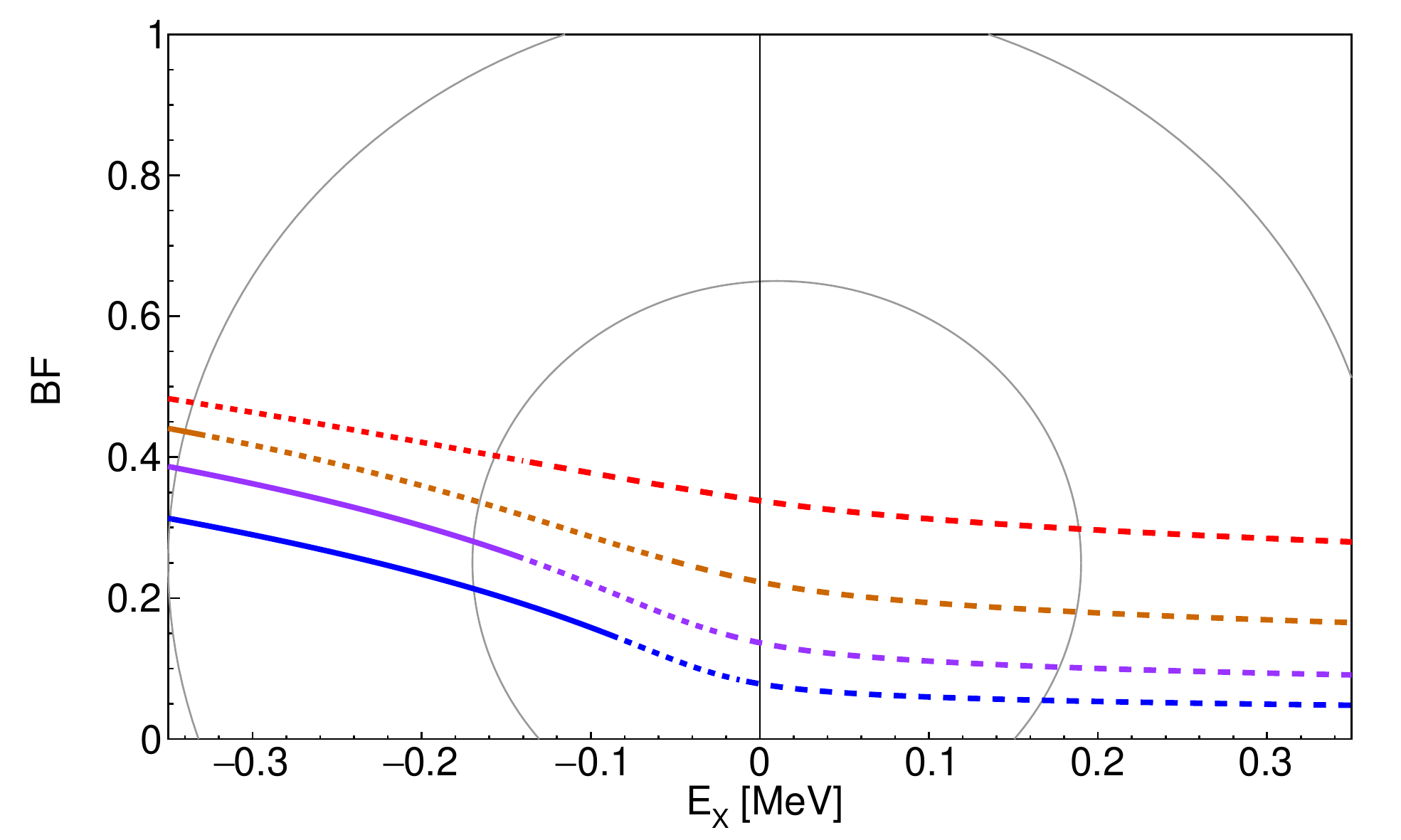} 
\caption{
Branching fraction  for short-distance decays of the  $X$ resonance feature 
as a function of the resonance energy $E_X$ for four values of $\mathrm{Im}[\gamma_X]/\sqrt{\mu \Gamma_{*0}}$:  
2, 1, 1/2, and 1/4  in order of decreasing branching fractions.
The branching fraction implied by Eq.~\eqref{BR-Xres} 
is shown as a curve that changes from solid for a narrow bound state
to dotted for a bound state that is not so narrow and then to dashed for a virtual state.
The thin grey curves are the $1\sigma$ and $2\sigma$ error ellipses for $E_X$ in Eq.~\eqref{EX-exp}
and the complement $(25\pm 40)\%$ of the branching fraction in Eq.~\eqref{BrDDpigamma}.
}
\label{fig:BFres}
\end{figure}
%%%%%%%%%%%%%%%%%%%%%%%%%%%%%%%%%%%%%%%%%%%%%%

If $E_\mathrm{min}$ and $E_\mathrm{max}$ are specified, Eqs.~\eqref{EX-def} and \eqref{BR-Xres} 
can be used to determine the two unknown parameters $\mathrm{Re}[\gamma_X]$ and $\mathrm{Im}[\gamma_X]$ 
given the measured  value of the resonance energy $E_X$ in Eq.~\eqref{EX-exp}
and the estimate of the CD branching fraction for the $X$ resonance feature in Eq.~\eqref{BrDDpigamma}.
Alternatively, for a given value of $\mathrm{Im}[\gamma_X]$,  Eq.~\eqref{EX-def}
can be used to determine $\mathrm{Re}[\gamma_X]$ as a function of $E_X$.
The SDD branching fraction BF for the $X$ resonance feature implied by Eq.~\eqref{BR-Xres} 
can then be predicted as a function of $E_X$.  
It can be compared to the estimated value $\mathrm{BF}= (25\pm 40)\%$ given by the complement of Eq.~\eqref{BrDDpigamma}.
In Fig.~\ref{fig:BFres}, the SDD branching fraction  BF for the $X$ resonance feature is shown 
as a function of the resonance energy $E_X$ for various values of $\mathrm{Im}[\gamma_X]$.
A curve changes from solid, indicating a narrow bound state, to dotted when $\mathrm{Re}[\gamma_X]$
decreases through the value in Eq.~\eqref{Regamma:nbs}.
The curve changes from dotted to dashed when $\mathrm{Re}[\gamma_X]$ becomes negative, 
indicating that the bound state has become a virtual state.
The error ellipses in Fig.~\ref{fig:BFres} are for the measured resonance energy $E_X$  in Eq.~\eqref{EX-exp} 
and for the estimated branching fraction $\mathrm{BF}= (25\pm 40)\%$.
The curves within the $1\sigma$ error ellipse are compatible with values of 
$\mathrm{Im}[\gamma_X]/\sqrt{\mu \Gamma_{*0}}$ up to about 9. 
The majority of the area inside the $1\sigma$ error ellipse corresponds to virtual states,
but there  is a region with negative $E_X$ that corresponds to narrow bound states.

%%%%%%%%%%%%%%%%%%%%%%%%%%%%%%%%%%%%%%%%%%%%%%%%
\begin{figure}[t]
\includegraphics*[width=0.8\linewidth]{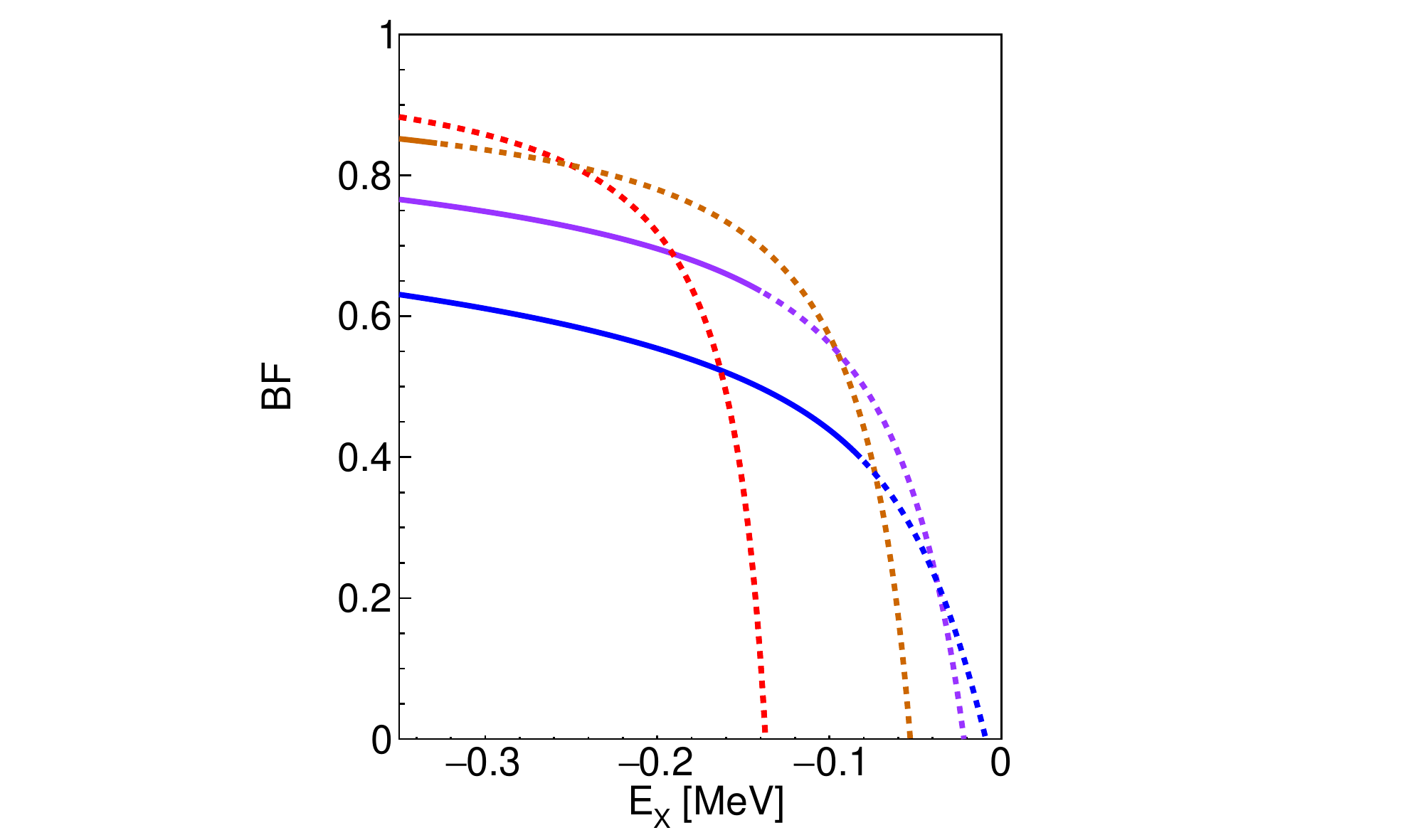} 
\caption{
Branching fraction for short-distance decays of the  $X$ bound state  as a function of the resonance energy $E_X$
for four values of $\mathrm{Im}[\gamma_X]/\sqrt{\mu \Gamma_{*0}}$:  
 2, 1, 1/2, and 1/4  in order of decreasing branching fractions on the left side.
The approximation for the branching fraction implied by Eq.~\eqref{BR-Xbound} 
is shown as a curve that changes from solid for a narrow bound state 
to dotted when the bound state is not so narrow.
}
\label{fig:BFbound}
\end{figure}
%%%%%%%%%%%%%%%%%%%%%%%%%%%%%%%%%%%%%%%%%%%%%%

The approximation in Eq.~\eqref{BR-Xbound} for the SDD over CD branching ratio of the  $X$ bound state
 implies a corresponding approximation for the SDD branching fraction of the  $X$ bound state.
In Fig.~\ref{fig:BFbound}, we show the  SDD branching fraction of the  $X$ bound state 
as a function of the resonance energy $E_X$ for various values of $\mathrm{Im}[\gamma_X]$.
A curve changes from solid, indicating a narrow bound state, to dotted
when $\mathrm{Re}[\gamma_X]$ decreases through the value in Eq.~\eqref{Regamma:nbs}.
The dotted curve decreases to zero when $\mathrm{Re}[\gamma_X]$ is 0,
indicating that the bound state is about to become  a virtual state.
In most of the region that corresponds to narrow bound states,
the branching fraction in Fig.~\ref{fig:BFbound} is considerably larger 
than the corresponding  branching fraction for the $X$ resonance feature in Fig.~\ref{fig:BFres}.
Within the $1\sigma$ error ellipse, the ratio of the branching fraction
for the $X$ bound state over the $X$ resonance feature ranges from 2.5 to 3.2.
The simplest model for the $X$ line shapes predicts that the
$J/\psi\, \pi^+\pi^-$ branching fraction of a narrow bound state should be larger than 
that of the $X$ resonance feature in Eq.~\eqref{BrXresJpsipipi} by that same factor.

%\newpage

%%%%%%%%%%%%%%%%%%%%%%%%%%%%%%%%%%%%
\section{Summary}
\label{sec:Summary}
%%%%%%%%%%%%%%%%%%%%%%%%%%%%%%%%%%%%

Preliminary measurements by the BaBar collaboration of the recoil momentum distribution 
of the $K^+$ in $B^+$ decays have been used to determine the $J/\psi\, \pi^+\pi^-$ branching fraction of the 
$X(3872)$ resonance feature from $B^+$-to-$K^+$ transitions in Eq.~\eqref{BrXresJpsipipi}.
We have emphasized that this is not a conventional branching fraction that must be independent of the production process.
If $X$ is a narrow bound state, its branching fractions must be the same for all short-distance production mechanisms,
 because they can be obtained by factoring amplitudes at the pole associated with the bound state.
However a branching fraction of the $X$ resonance feature is obtained by integrating 
over the energy of the resonance feature, which includes a threshold enhancement peak 
as well as a possible bound-state peak,
and it should therefore be expected to depend on the production mechanism.
We used a previous measurement by the Belle collaboration to
obtain the $D^0 \bar D^0 \pi^0$ branching fraction of the $X$ resonance feature from $B^+$-to-$K^+$ transitions 
in Eq.~\eqref{BRDD*Belle}.
A more precise measurement of the $D^0 \bar D^0 \pi^0$ branching fraction of the $X$ resonance feature  
and a measurement of its $D^0 \bar D^0 \gamma$ branching fraction would be useful.

The current experimental result for the $X$ resonance energy $E_X$, which is obtained from measurements 
in the $J/\psi\, \pi^+\pi^-$ decay mode, is given in Eq.~\eqref{EX-exp}, and it is rather precise.  
We introduced a theoretical prescription for the $X$ resonance energy in Eq.~\eqref{Eres-def} 
that can be useful for constraining the  parameters of models for the $X$ line shapes.
This prescription  is just the center of energy in the $i=J/\psi\, \pi^+\pi^-$ decay mode.
We illustrated it by applying it to the simplest plausible model for the $X$ line shapes,
which can be specified by the scattering amplitude in Eq.~\eqref{Amp-E}
and by the energy range  from $E_\mathrm{min}$ to $E_\mathrm{max}$.
We derived analytic approximations for $E_X$ in three limits:
the bound-state limit in Eq.~\eqref{EX-bound},
the zero-energy resonance in Eq.~\eqref{EXzero},
and the virtual-state limit in Eq.~\eqref{EXvirtual}.

Our estimate of the branching fraction into constituent decay modes of the $X$ resonance feature 
from $B^+$-to-$K^+$ transitions in Eq.~\eqref{BrDDpigamma} implies that the complementary branching fraction  
into short-distance decay modes is $\mathrm{BF} = (25 \pm40)\%$.
Using the measured value of $E_X$ in  Eq.~\eqref{EX-exp} and our estimate of BF, 
we obtained constraints on the parameters of the simplest plausible model for the $X$ line shapes.
Further measurements of branching fractions of $X$ resonance features could be used to determine the constraints 
on the parameters for more realistic models of the $X$ line shapes that explicitly take into account additional channels, 
such as the charged-charm-meson-pair channel  or the $\chi_{c1}(2P)$ channel.

If the $X$ is a narrow bound state,
the branching fraction of the $X$ bound state  into a short-distance decay mode
can be substantially larger than the corresponding branching fraction of the $X$ resonance feature.
The $J/\psi\, \pi^+\pi^-$ branching fraction of the $X$ bound state  can therefore be substantially larger than 
the $J/\psi\, \pi^+\pi^-$ branching fraction of the $X$ resonance feature.
A loose lower bound on the $J/\psi\, \pi^+\pi^-$ branching fraction of the $X$ bound state 
is provided by the $J/\psi\, \pi^+\pi^-$ branching fraction of the $X$ resonance feature in Eq.~\eqref{BrXresJpsipipi}.
An upper bound on the $J/\psi\, \pi^+\pi^-$ branching fraction of the $X$ bound state 
is given in Eq.~\eqref{BrJpsipipi<}.

The $X$ can be produced by the creation of a $D^*\bar D^*$ pair at short distances
followed by the rescattering of the charm mesons into $X\pi$ or $X\gamma$.
A charm-meson triangle singularity produces a narrow peak in the invariant mass distribution for $X\pi$ or $X\gamma$ 
near the  $D^*\bar D^*$ threshold  \cite{Guo:2019qcn,Braaten:2019gfj}.
Under the assumption that $X$ is a narrow bound state,
we have calculated the production rate of $X\pi$ near the peak from the triangle singularity 
in  hadron colliders \cite{Braaten:2018eov,Braaten:2019gfj} and in $B$ meson decay \cite{Braaten:2019yua}.
We also calculated the cross section for electron-positron annihilation into $X \gamma$ near the 
peak from the triangle singularity \cite{Braaten:2019gfj}.
For the coupling of the $X$ bound state to the charm-meson pairs, we used a vertex determined by the binding energy. 
The cross section for producing $X$ in the specific decay mode $J/\psi\, \pi^+\pi^-$
is obtained by multiplying the cross section for producing the $X$ bound state 
by the $J/\psi\, \pi^+\pi^-$ branching fraction for the $X$ bound state.
That branching fraction should be considerably 
larger than the roughly 4\% branching fraction of the $X$ resonance feature  in Eq.~\eqref{BrXresJpsipipi},
and it should be smaller than the upper bound of 33\% in Eq.~\eqref{BrJpsipipi<}.
Tighter constraints on this branching fraction would allow more precise  predictions of the heights of the 
peaks from the triangle singularity observed through the $J/\psi\, \pi^+\pi^-$ decay mode of the $X$.
The observation of these peaks would provide convincing evidence 
for the identification of the $X$ as a weakly-bound charm-meson molecule.

{\it Note added:}
As this paper was being finalized,
Li and Yuan posted a paper entitled
``Determination of the absolute branching fractions of $X(3872)$ decays'' \cite{Li:2019kpj}.
They presented a complete analysis of all the existing data that can give $X$ branching fractions.
Their branching fractions should not be interpreted as those for the $X$ bound state,
because their inputs included several results for $X$ resonance features from various production mechanisms.
In addition to the inclusive branching fraction of  $B^+$ into $K^+$ plus the $X$ resonance feature from Ref.~\cite{Wormser},
they included results for the $D^{*0} \bar D^0$ and $D^0 \bar D^{*0} $ final states from
the $X$ resonance feature from $B^+$-to-$K^+$ transitions, from $B^0$-to-$K^0$ transitions,
and from $e^+ e^-$-to-$\gamma$ transitions.
Their analysis relied on the unjustified assumption that the branching fractions for $X$ resonance features
are the same for these three production mechanisms.

%\newpage 

%\appendix

%%%%%%%%%%%%%%%%%%%%%%%%%%%%%%%%%%%%%%%%%%
\begin{acknowledgments}
This work was supported in part by the Department of Energy under grant DE-SC0011726
and by the National Science Foundation under grant PHY- 1607190.
We thank F.K.\ Guo, E.\ Johnson, and C.Z.\  Yuan for useful comments.
\end{acknowledgments}
%%%%%%%%%%%%%%%%%%%%%%%%%%%%%%%%%%%%%%%%%%

%%%%%%%%%%%%%%%%%%%%%%%%%%%%%%%%%%%%%%%%%%

 \end{document}